\documentclass[times,authoryear]{elsarticle}

\usepackage{jasr}
\usepackage{framed,multirow}
\usepackage{commons}
\usepackage{gensymb}

\usepackage{amssymb}
\usepackage{latexsym}

\usepackage[switch]{lineno}

\usepackage{url}
\usepackage{xcolor}
\definecolor{newcolor}{rgb}{.8,.349,.1}

\usepackage[colorlinks, citebordercolor=white]{hyperref}

\usepackage{siunitx}
\usepackage{cleveref}
\usepackage{bm}

\DeclareMathOperator*{\argmin}{arg\,min}

\journal{Advances in Space Research}

\begin{document}

\verso{Stephen Catsamas \textit{et al.}}

\begin{frontmatter}

\title{Thermal infrared characterization of spatially unresolved resident space objects: Prospects from analytical two-component modeling}

\author[1]{Stephen \snm{Catsamas}\corref{cor1}}
\cortext[cor1]{Corresponding author: 
  email: scatsamas@student.unimelb.edu.au}
    \author[1]{Sarah \snm{Caddy}}
  \author[1]{Michele \snm{Trenti}}
\author[1]{Benjamin \snm{Metha}}
\author[1]{Simon \snm{Barraclough}}
\author[1,2]{Robert \snm{Mearns}}
\author[2]{Airlie \snm{Chapman}}
\author[1]{Rachel \snm{Webster}}

\affiliation[1]{organization={School of Physics, The University of Melbourne},
                addressline={Grattan Street},
                city={Parkville},
                postcode={3010},
                country={Australia}}

\affiliation[2]{organization={Faculty of Engineering and Information Technology, The University of Melbourne},
                addressline={Grattan Street},
                city={Parkville},
                postcode={3010},
                country={Australia}}


\begin{abstract}

In this work we investigate the potential of a thermal infrared (IR) space telescope to remotely characterize the component temperatures of a satellite.  With the rapid increase in the number of objects launched in recent years, the ability to detect, track, identify and determine the intent of satellites has become of increasing importance. Spectral modeling of satellites from multi-wavelength photometry in the thermal IR is a technique that has the potential to derive information about the temperature and operational status of a satellite in orbit, without the requirement to spatially resolve the target. Previous work has focused on determination of a single/effective temperature for a Resident Space Objects (RSOs) -- such as satellites, asteroids, debris and rocket bodies -- from remote observations, obtaining mixed results in terms of ability to classify objects. 
To progress, we explore a two-greybody component spectral model. Using this analytical model, we investigate which temperature characteristics may be identified from unresolved multi-wavelength photometric observations as a function of the signal-to-noise ratio, under the assumption of Poisson noise-dominated data. With this instrument-agnostic framework, we then quantify the potential of this model to discriminate between RSOs with a single temperature (e.g. natural rocks) versus human-made satellites with a chassis and deployed solar panels where significant component temperature differences exist under typical orbital configurations. Last, we comment on promising prospects of this model for applications to existing and future space telescope observations to characterize RSOs from spatially unresolved photometry.
\end{abstract}


\end{frontmatter}


\section{Introduction} \label{sec:intro}

\begin{figure}
    \centering
    \includegraphics[]{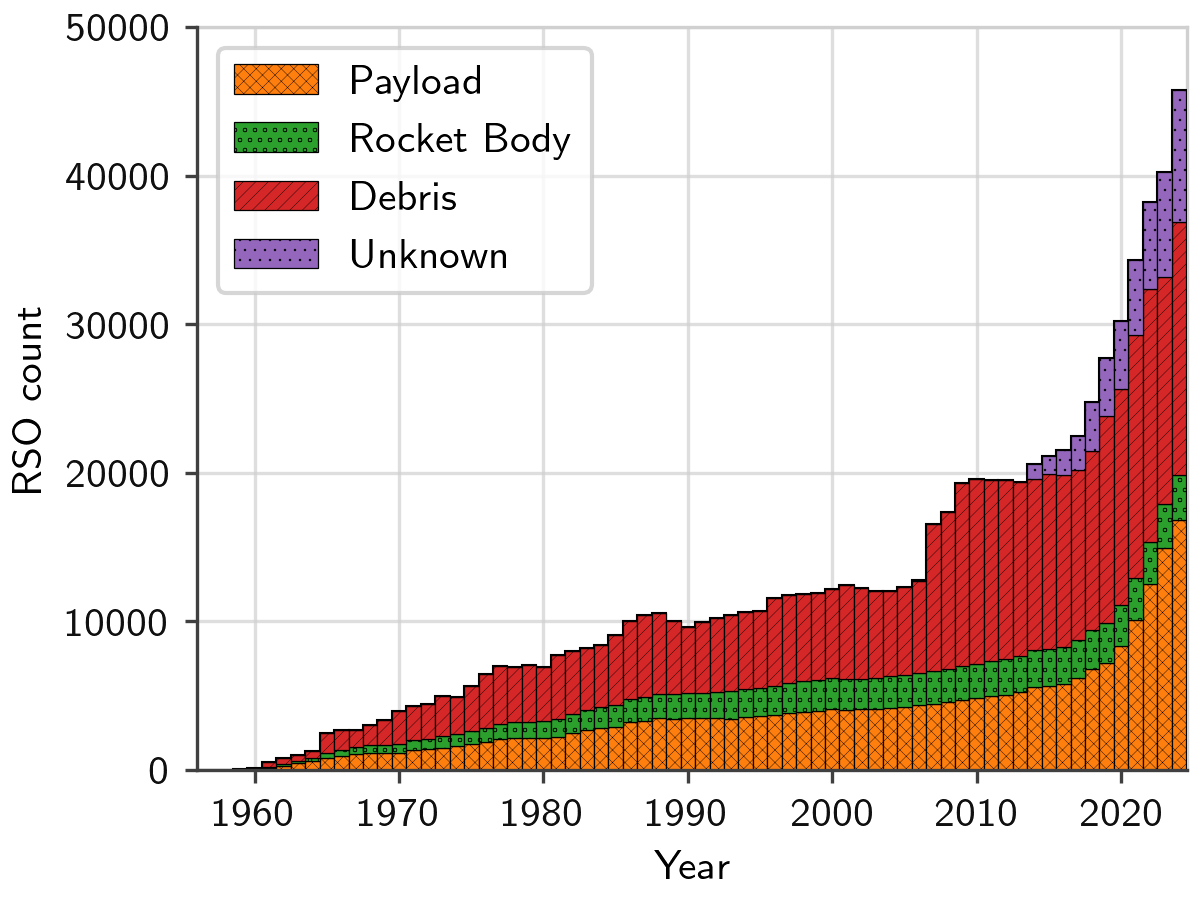}
    \caption{Trackable RSO count evolution over time by classification. The number of objects in orbit has more than doubled in the last 10 years, and about \qty{22}{\percent} of current objects have unknown classification. Extrapolating, this trend of a larger number of satellites in orbit will require investment in new technology to expand existing SDA networks. In addition, the demand for more sophisticated space surveillance to ensure the security of critical infrastructure also highlights the need to broaden the capabilities of SDA facilities. Data sourced from ESA DISCOS (Database and Information System Characterising Objects in Space; 2024).}
    \label{fig:sat_launch_rate}
\end{figure}

Space is becoming increasingly congested and contested \citep{bloom_space_2022}. In the past decade alone, the rate of increase in the number of Resident Space Objects (RSOs) in orbit has more than tripled (\cref{fig:sat_launch_rate}). As a result, the ability to determine the location, identity, operational status and intent of RSOs has become critical for space sustainability efforts \citep{commonwealth_of_australia_2020_2020, united_nations_office_for_outer_space_affairs_guidelines_2021, sejba_space_2023}. Capabilities used to inform active decision making for space users are known as Space Domain Awareness (SDA) (see, e.g., \citealt{bloom_space_2022}).

One method of gathering intelligence of RSO's is with remote thermal observations. Typically, previous works have considered modelling the thermal signature of an RSO assuming a single characteristic temperature from an unresolved point source \citep{skinner_ir_2007, skinner_observations_2014, paxson_space_2008, wright_infrared_2023}. However, this simplification results in information being lost when the RSO is a complex multi-component spacecraft. In this work, we explore the use of a spectral modelling method that may be used in future observational studies to derive information such as power flows, thruster firing and spacecraft health from a more complex two component thermal model. We present an instrument agnostic framework, considering theoretical Poisson noise dominated observations with no introduced systematic uncertainty, to determine the suitability of this methodology to produce meaningful component temperature estimates within these assumptions.

The field of space domain awareness encompasses a range of techniques which are used to collect different types of intelligence of RSO's. To motivate this work, we present a broad overview of the methods currently used to monitor RSO's including their benefits and limitations.

\subsection{Ground-based Space Surveillance Systems}
The majority of activities to develop SDA capability have been focused on using a combination of ground-based radar and electro-optical detectors \citep{sejba_space_2023}. Global SDA networks, such as The United States Space Surveillance Network, consist of a large number of such detectors located around the globe and are designed for the purpose of comprehensive detection, identification, tracking and cataloging of all human-made objects orbiting the Earth \citep{johnson_us_1993, chatters_space_2009}. Observations from the United States Space Surveillance Network are combined into a catalog of RSOs
\citep{miller_new_2007}\footnote{Access to the Space Track catalog is available online
  at \url{space-track.org}.} which can be used to notify owners
of satellites about potential collision risks, in order to protect space
assets. 
\\
\\
While ground based sensors have played an important role in SDA operation to date, they do have some limitations. Ground based optical systems are limited to observing satellites when they are lit by an external light source (usually direct sunlight, or reflected by the Earth or Moon). They are also dependent on the weather, and require clear atmospheric conditions to operate effectively. The development of daytime optical SDA is improving the operational constrains of some ground based optical systems despite the intrinsically higher background noise levels (see e.g. \citealt{shaddix_daytime_2021, caddy_surprising_2024, caddy_optical_2024}), however many facilities are still limited to operating at night, impacting the latency of observations. Furthermore, single optical SDA facilities are geographically limited, and are only able to observe objects in Low Earth Orbit (LEO) when they pass overhead, and have a limited observational horizon from which to observe geosynchronous orbit (GEO) satellites at fixed longitudes. 
\\
\\
Passive and active radar facilities benefit from being able to observe in any atmospheric conditions and are able to observe multiple objects over large swathes of the sky quickly. However, these facilities have different challenges for monitoring RSOs. Given that active radar is based on detection of the reflected signal it emitted, the signal strength falls with the fourth power of distance. Therefore, there are a limited number of facilities that are powerful enough to be capable of monitoring objects past the edge of LEO at \qty{2000}{\kilo \meter} and reach large GEO satellites (e.g., Norway's GLOBUS II, see \citealt{weeden_global_2001}). Furthermore, concepts for "stealth" satellites, where the radar cross section is minimized, have been investigated theoretically and possibly demonstrated in orbit \citep{stealth_sat_2022}. Passive radar systems do not produce their own signal, and rely on detecting a signal emitted from a target during nominal operations. One of the key disadvantages of these systems is that they cannot track objects that do not actively emit, such as space debris, rocket bodies, natural hazards like asteroids and inoperable satellites \citep{sejba_space_2023}.     
\\
\\
The demand for deep space SDA will only increase in the next decade as cislunar activities become more frequent and small satellites (which are harder to detect because of their smaller geometric cross-section) become more common. Missions such as NASA's Artemis and Lunar Gateway \citep{block_cislunar_2022}, as well as the advent of lunar operations by private companies like the robotic lunar lander of Intuitive Machines\footnote{\url{https://nssdc.gsfc.nasa.gov/nmc/spacecraft/display.action?id=IM-1-NOVA}} and Astrobotic \footnote{\url{https://nssdc.gsfc.nasa.gov/nmc/spacecraft/display.action?id=PEREGRN-1}} will place greater demand on existing SDA facilities in a larger volume of space. Challenges include the identification of unresolved objects, and precise orbit determination in a complex 3 body gravitational potential field (the Earth, Sun and Moon; e.g. see~\citealt{baker_comprehensive_2024}).

\subsection{Space-based Space Surveillance Systems}
Space-based SDA systems are one solution that may be used to improve and extend the capabilities of SDA networks \citep{utzmann_space-based_2014, du_tentative_2019}. Space-based SDA is currently being explored across multiple government organizations, as well as by private companies \citep{gaposchkin_space-based_2000, leitch_sapphire_2010, ackermann_systematic_2015, allworth_use_2024, oconnell_concept_2024}. Multiple government owned facilities have also been designed and flown in recent years to take advantage of the benefits of space-based SDA systems \citep{gaposchkin_space-based_2000, leitch_sapphire_2010, utzmann_space-based_2014, ackermann_systematic_2015, du_tentative_2019}. The U.S. Air Force Space Surveillance Network's Space-based Surveillance System (SBSS) \citep{utzmann_space-based_2014} is a \qty{30}{\centi \meter} pathfinder space telescope designed for SDA and was launched in 2010. The system is capable of both actively tracking satellites, as well as sidereal tracking (where background stars remain stationary) to determine precise orbital elements for the observed targets \citep{ackermann_systematic_2015}. This pathfinder has since been expanded upon to become a constellation of satellites including Sapphire \citep{leitch_sapphire_2010}, Block 10, GSSAP\footnote{\url{https://www.spaceforce.mil/About-Us/Fact-Sheets/Article/2197772/geosynchronous-space-situational-awareness-program/}} and NEOSSat \citep{du_tentative_2019}. 
\\
\\
Small satellites like Sapphire and NEOSSat play an important role in space-based SDA, with fast development times and lower costs. For example, the Sapphire satellite is a space telescope with a \qty{15}{\centi \meter} aperture operated for the United States Space Surveillance Network by the Canadian government \citep{leitch_sapphire_2010}. While the system is less productive than SBSS (capable of 1600 observations per day compared to SBSS's 21000) the mission achieved its goals for a smaller cost and continues to contribute to the Space Surveillance Network \citep{ackermann_systematic_2015}, demonstrating the benefits of specialized small satellites for space-based SDA. 
\\
\\
One novel area of active development is Non-Earth Imaging (NEI) - the use of satellites (typically constellations) with optical imaging capabilities that perform high resolution imagery of RSOs in LEO. Commercial constellations of space-based sensors designed for fast cadence, high resolution imaging of RSOs up to orbit heights of \qty{750}{\kilo \meter} observed in favorable conditions during flybys are emerging \citep{allworth_use_2024,oconnell_concept_2024}. These systems are able to achieve diffraction-limited imaging in orbit. The data quality is not negatively impacted by the atmosphere in regards to photon absorption (cloud cover), blurring induced by turbulence, or elevated background from scatter during daytime. 
\\
\\
However, even NEI resolution is intrinsically limited by the distance between the telescope and the target, hence the performance is rapidly degrading if favorable close encounters between telescope and target in LEO are not available. It is not possible to resolve GEO targets even with the largest space-based systems in LEO (and NEI from a GEO telescope would require a rare favorable conjunction). In fact, considering the ideal (optimal) case of a circular-aperture diffraction-limited system,  a \qty{40}{\centi \meter} aperture space telescope in LEO observing in blue light (\qty{0.4}{\micro \meter}) is incapable of resolving details smaller than $\sim$ \qty{40}{\meter} for a target in GEO, therefore providing essentially unresolved photometric information only. Even the largest space-based optical telescopes like the Hubble Space Telescope, with a \qty{2.4}{\meter} aperture, cannot resolve details smaller than $\sim$ \qty{7}{\meter} in GEO. 

\subsection{Thermal Infrared Space Domain Awareness}

\begin{figure}
    \centering
    {\includegraphics[width=1.0\textwidth]{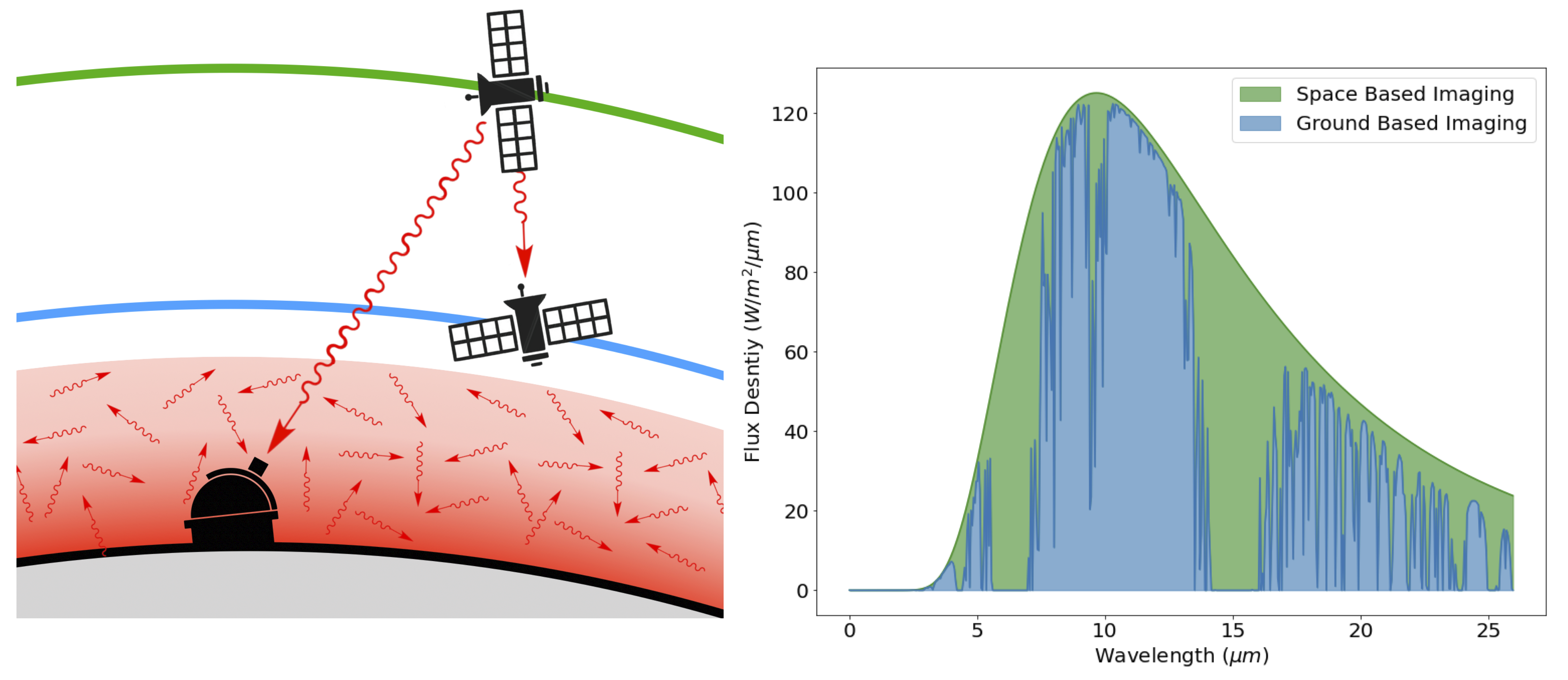}}
    \caption{Left panel: Schematic of a SBSS system at IR wavelengths. From its
      vantage point in low Earth orbit, a space telescope will be able
      to measure temperature and power fluctuations of other
      satellites by means of high signal-to-noise infrared
      observations of their thermal signatures. Such inference would
      not be possible from ground-based observatories as infrared
      observations are affected from substantial foreground noise
      emitted by the molecules in the atmosphere, illustrated in the right panel.}
    \label{fig:schematic}
\end{figure}

\begin{figure}
    \centering
    \includegraphics[scale=1.0]{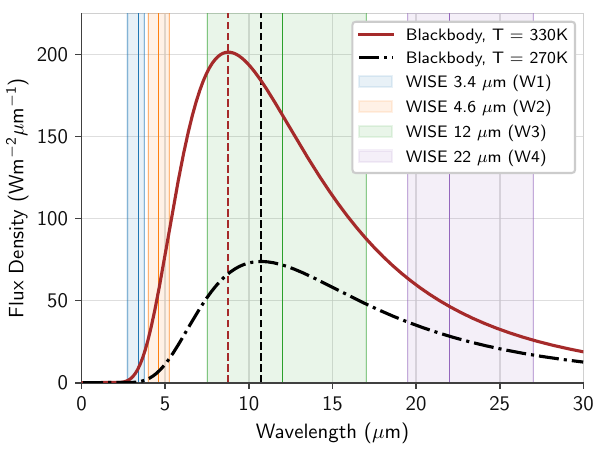}
    \caption{
      WISE space telescope bandpasses shown as a function of wavelength, with the shaded region providing the width of the bandpass and the solid line providing the band's central wavelength.
      The bands are shown in blue for W1 at \qty{3.4}{\micro \meter}, orange for W2 at \qty{4.6}{\micro \meter}, green for W3 at \qty{12}{\micro \meter}, and purple for W4 at \qty{22}{\micro \meter}. The maroon and black series show blackbody spectra at \qty{330}{\kelvin} ($\approx$ \qty{57}{\degreeCelsius}) and \qty{270}{\kelvin} ($\approx$ \qty{-4}{\degreeCelsius}) respectively. The WISE bands sample the peak of the distribution. Observations in similar bands from the Midcourse Space Experiment satellite have previously been used to determine the temperature of RSOs \citep{paxson_space_2008}. As such, this work considers these bands' central wavelength as the baseline for investigating a more sophisticated two-component thermal model.
      }
    \label{fig:wise_bandpasses}
\end{figure}

Thermal infrared analysis of RSOs is possible because all objects reject heat by radiation to reach a thermal steady state whereby this radiated energy is balanced by sources such as the incident energy from the Sun or energy stored in internal batteries. Given that the steady state temperature of a spherical blackbody at one astronomical unit from the Sun is about \qty{280}{\kelvin}, the emitted thermal spectrum of RSOs peaks in the infrared wavelengths at roughly \qty{10}{\micro \meter}. An important consequence of this is that even satellites designed to be masked from optical and/or radar detection must still radiate in the IR to reject heat and thus may be visible to thermal infrared sensors.
Infrared space telescopes such as WISE have appropriate bandpasses to fit a greybody spectrum at the expected RSO temperatures. This concept is illustrated in \cref{fig:wise_bandpasses}.
\\
\\
Differing types of RSOs exhibit distinct thermal behaviors which may facilitate their characterization. For example, natural rocks of radius $> \qty{1}{\meter}$ have a large heat capacity \citep{ceplecha_meteor_1998} and will remain at a near-constant temperature, even during eclipse periods. Space debris will have no active thermal control and therefore will have a temperature determined by insolation, orientation, geometry, and surface emissivity and absorptivity. Sun-pointing solar panels on an operational satellite will have a constant (but seasonally varying) insolation during non-eclipse periods and thus their temperature will largely be determined by the amount of power drawn from them \citep{kim_analytical_2010,gilmore_spacecraft_2002}. Solar panels are expected to be cooler when used to power intensive operations such as battery charging, on-board data processing, data transmission, or orbital maneuvering. Conversely, solar panels are expected to be warmer while idle. Furthermore, using data from \citet{liu_dynamic_2019}, the low thermal mass of solar panels mean that these temperature shifts are rapid - initially on the order of \qty{1}{\kelvin \per \minute} when transitioning from zero to full power draw and on the order of \qty{10}{\kelvin \per \minute} upon entering eclipse. If the temperature of an RSO can be determined accurately from remote observations, frequent monitoring may be used to discriminate between both different RSO types and infer onboard operations of active RSOs.
\\
\\
To date, traditional SDA facilities have predominantly focused on optical and radar observations as a method of localizing, monitoring and characterizing RSOs. However, these techniques suffer limitations which result in current SDA facilities being ineffective in the optical when an object is in shadow (e.g. night side of LEO orbit or eclipsed in GEO). At radar wavelengths, limitations occur when an object is not emitting (for passive radar), or is too small or too far away to been seen (for active radar). Thermal infrared observations ($\lambda \approx \qtyrange{1}{100}{\micro \meter}$) are an emerging capability which addresses the shortcomings of traditional SDA facilities \citep{wright_infrared_2023, skinner_observations_2014, paxson_space_2008}. Thermal infrared observations detect the thermal emission spectrum of a satellite. These emissions result from rejecting heat from either solar radiation or heat produced under the satellite's own power. 
As the emission is determined primarily by the satellite's own temperature, it does not rely on the object actively reflecting light from an external source. 
As a result the thermal emission spectrum of an RSO can be monitored during eclipse, when the target is poorly illuminated and challenging to observe at visible wavelengths \citep{wright_infrared_2023}.
\\
\\
Interest in thermal infrared observations of RSOs has grown in recent years to include space-based observations taken by astronomical space telescopes such as NASA's Wide-field Infrared Survey Explorer (WISE) \citep{lee_infrared_2016, lee_distinguishing_2017, fitzmaurice_streak_2018, fitzmaurice_detection_2021, wright_infrared_2023} and Infrared Astronomical Satellite (IRAS) \citep{dow_detection_1990, gaposchkin_infrared_1995} as well as dedicated space-based surveillance systems such as the Midcourse Space Experiment (MSX) \citep{gaposchkin_space-based_2000, paxson_space_2008}. In addition, interest has also grown in developing ground based thermal infrared capability including taking spectra of satellites using large ($> \qty{1}{\meter}$ class) telescopes for the purpose of identification and monitoring of their thermal properties \citep{skinner_ir_2007, gibson_optical-infrared_2013, skinner_observations_2014, hart_quantitative_2014}. Detection and photometry of satellites in the short wave infrared (\qtyrange{0.9}{2}{\micro \meter}) has also been attempted from the ground \citep{hart_resolved_2015, thomas_ground-based_2019, pearce_rapid_2021, kielkopf_remote_2022} as well as analytical studies demonstrating the benefits of thermal infrared photometry for identification of the material properties of satellites \citep{reyes_analysis_2023}. We discuss the development of these capabilities in more detail below.
\\
\\
Monitoring the temperature of RSOs (either for identification or in the case of satellites for operational monitoring) is a key advantage of the use of thermal infrared observations for space-based SDA. Thermal modeling for RSOs such as space debris has been studied \citep{mccall_thermal_2013} and thermal infrared observations have already shown promise in determining the temperature of RSOs detected from the ground through IR atmospheric windows. For example, observations of 10 GEO satellites have been conducted from \qtyrange{1}{2.5}{\micro \meter} using the Hi-VIS Echelle spectrograph, and the Broadband Array Spectrograph System from \qtyrange{3}{13}{\micro \meter} on the Advanced Electro-Optical System (AEOS) 3.6 meter telescope at the Air Force Maui Optical and Supercomputing site \citep{skinner_ir_2007}. They report between \qtyrange{4}{10}{\percent} photometric accuracy of their targets and observe satellites at a range of solar phase angles (the angle between the Sun, satellite and observer). Using a single greybody thermal model, temperatures estimates are derived between \qtyrange{320}{490}{\kelvin} ($\approx$ \qtyrange{40}{200}{\degreeCelsius}), with warmer temperatures occurring at phase angles of \qty{90}{\degree}, and are speculated to be attributed to the emission being dominated by hot, Earth facing transmitters \citep{skinner_ir_2007}. In addition, AEOS was used to demonstrate observations of a satellite in both active station keeping operations, as well as again after decommissioning, and found cooler temperatures between \qtyrange{260}{320}{\kelvin} ($\approx$ \qtyrange{-13}{40}{\degreeCelsius}) \citep{skinner_observations_2014}. Observations were noted to not adhere to the phase angle relationship which is attributed to uncontrolled tumbling \citep{skinner_observations_2014}. Furthermore, IR broadband color observations have the potential to discern between reflectance spectra of different material properties found on satellites \citep{reyes_analysis_2023}. This technique has already been demonstrated as a promising method of discriminating between material types on a satellite, and used for identification \citep{pearce_rapid_2021}.
\\
\\
Expanding SDA capabilities to longer IR wavelengths from space has the potential to provide new insight into the operational state and identity of RSOs \citep{paxson_space_2008, fitzmaurice_streak_2018, wright_infrared_2023}. Infrared observations from ground-based telescope facilities must contend with the infrared sky background (see e.g. \citealt{cox_infrared_2002} for a comprehensive review). They are restricted to observing in a limited number of ``atmospheric windows'' where the atmosphere is semi-transparent to infrared radiation due to absorption/emission from water vapor, OH and CO2 gases \citep{ellis_case_2008}. This restricts ground based IR observations to a handful of sites with optimal atmospheric conditions, such as NASA's Infrared Telescope Facility on Mauna Kea in Hawaii and the Visible and Infrared Survey Telescope for Astronomy (VISTA) in the Atacama desert in Chile, where humidity is low. Even in the best sites, the IR sky background is still variable \citep{glasse_infrared_1993}. In contrast, these complications do not affect space-based data collection giving such an advantage that --- at mid-infrared wavelengths --- small aperture space telescopes (e.g. NASA's Spitzer and WISE space telescopes) outperform the best facilities on the ground \citep{gehrz_nasa_2007}. A visualization of this concept is provided in \cref{fig:schematic}.
\\
\\
Thus, in-principle, space-based IR SDA systems offer the opportunity for a new capability: high precision IR photometry and spectroscopy for satellite identification and operational monitoring. IR space-based SDA has been attempted from dedicated satellites, as well as serendipitous observations from astronomical space telescopes. The NASA WISE space telescope is one such instrument that has been used to investigate the potential for space-based IR SDA  \citep{howard_geo_2015, lee_infrared_2016, lee_distinguishing_2017, fitzmaurice_streak_2018, fitzmaurice_detection_2021}. WISE's primary mission was the mapping of the entire infrared sky in 4 broadband filters centered on \qtylist{3.4; 4.6; 12; 22}{\micro \meter} \citep{wright_wide-field_2010}, improving upon the sensitivity and resolution of previous surveys conducted by the IRAS \citep{neugebauer_infrared_1984} and the AKARI mission \citep{murakami_infrared_2008}. Color distributions of RSOs detected in WISE catalogs have been demonstrated as a potential tool for distinguishing between satellites of different types, namely active box-wing type satellites, and cylindrical geostationary satellites \citep{lee_infrared_2016, lee_distinguishing_2017}. However, further analysis found that this technique is limited when considering satellite platforms of similar shape/type, which could not be differentiated using WISE color observations alone \citep{fitzmaurice_streak_2018, fitzmaurice_detection_2021}. In addition, a general linear relationship between size of the satellite and brightness could be established using WISE data \citep{fitzmaurice_streak_2018}. IRAS observations of RSOs have also been analyzed for the detection of sources \citep{dow_detection_1990} and photometry of RSOs to estimate object temperatures \citep{gaposchkin_infrared_1995}. In total, IRAS detected 190,000 RSOs, and 2047 were attributed to known catalogs of sources, enbling the estimation of the temperature, emissivity and absorptivity of these RSOs \citep{fitzmaurice_streak_2018}. A dedicated SDA space telescope, the Midcourse Space Experiment satellite, has also been used to investigate the temperature of RSOs in orbit \citep{paxson_space_2008}. RSOs observed included payloads, rocket bodies, and debris. Objects in eclipse (not directly illuminated by the Sun) were found to be on average cooler than objects in direct sunlight, however no distinction between the temperature of payloads and rocket bodies could be made from the observations \citep{paxson_space_2008}.

%
In this paper, we take a new approach to the challenge of spectral modelling for RSO identification and characterization by considering multiple RSO components, expanding upon current monolithic modelling approach of IR SDA. We focus on proof-of-principle and present the expected uncertainties of inferred spectral model parameters, such as component temperatures, to analyze the prospects of our approach for RSO identification and characterization.
\\ 
\\
This paper is organized as follows. \Cref{sec:concept} introduces the idea of IR thermal signature modeling for SDA with an approach inspired by astronomical modeling of galaxy spectra as a combination of blackbodies. \Cref{sec:thermal} develops a simple thermal model for a geostationary satellite, to provide the framework for a quantitative analysis. \Cref{sec:theory} presents our spectral fitting framework, and \Cref{sec:numerics} investigates the ability to measure the basic parameters of the simple thermal model by sampling the IR spectral energy distribution (SED) at multiple wavelengths. Results are discussed in \Cref{sec:discussion}, and \Cref{sec:conclusion} summarizes and concludes.

\section{Concept overview}\label{sec:concept}
The potential of spatially-unresolved SED modeling and observations for inferring properties of the object emitting the radiation is well illustrated in the context of astronomical observations of galaxies. The continuum emission of stars is, to first approximation, close to blackbody radiation, with the characteristic temperature and lifetime of stars depending on mass \citep{StellarEv_1967ARAA,HR_langer_2024}. Hence, the resulting emission is a linear combination of individual stellar spectra weighted by current distribution of stellar types, which in turn depends on the star formation history of the galaxy. Further, the spectrum is affected by corrections due to absorption of photons both in the stellar atmospheres and from the dust and gas present in the galaxies. Detailed modeling has been developed (e.g., \citealt{bruzual_stellar_2003}) and used for inferring galaxy properties such as stellar mass, characteristic stellar age, and dust/gas content from a small number of photometric observations in broad bands centered around the peak of the blackbody emission for stars, i.e. in the visible and near-IR range \citep{Johnson2021}. With spectroscopic data, it is even possible to reconstruct in detail the full star formation history of a galaxy \citep{Panter2007}.
\\
\\
These astronomical methods can be applied to SDA, as the SED of an RSO is the sum of the contributions from each of the emitting surfaces. Rather than modeling the spectrum from each emitting surface as a blackbody as done for a star, the spectrum from each emitting surface is better described by a greybody at the surface's temperature and effective area that depends on surface area, emissivity and geometric view factor to the observer. 
Therefore, by observing at different wavelengths, it should be in-principle possible to recover the temperatures and effective areas of the surface elements of the RSO without having to spatially resolve the target. The quality of the model that can be fit will depend on the number and placement of photometric bands, with increasing fidelity for greater spectral resolution. We predict that a small number of narrow IR bands will provide an estimate of one or two characteristic temperatures of the RSO. Consequently, an instrument designed with a larger number of band passes would likely be able to identify the presence of a greater number of distinct emitting surfaces. 
\\
\\
Degeneracies in parameter estimation affect SED modeling in astronomy, with combinations of different dust content, stellar ages and masses offering similar photometric solutions in some cases  \citep{brammer2008}. Naturally, these degeneracies need to be carefully assessed for RSOs and will likely introduce correlation between temperatures and effective emitting areas. However, the RSO observations have a key advantage. The geometry of an RSO and its surfaces are generally fixed, with the exception of deployable structures. Therefore, repeated observations of the same RSO from different viewing angles can offer the opportunity to build a higher fidelity model where degeneracies may be broken. Furthermore, if the RSO is in geostationary orbit, multiple observations could be carried out from low-Earth orbit at the same time by a constellation of space telescopes. Each element of the constellation would observe the RSO with a different viewing angle and enable a reconstruction of a 3 dimensional structure for the RSO. 
\\
\\
As a first step to assess potential of the new method, this paper introduces the concept with a two-component thermal model, which is used to forecast the inferred parameter uncertainties from sampling an SED at a limited number of wavelengths. The purposes of this paper are: 1) to demonstrate feasibility of this novel thermal inference method, 2) to characterize the dependence of its performance on factors such as component temperature difference and number of wavelength samples, and 3) to compare its inference against the simpler one-component thermal model which is generally accepted in the literature. We present this work as a foundational step for developing novel SDA capabilities (akin to foundational work for other SDA capabilities, e.g. \cite{cooke_classification_2025, cook_augmented_2025, mccall_thermal_2013}), and it will serve to guide future analysis of IR observations of RSOs.

\begin{figure}
\centering
\includegraphics[width=0.7\textwidth]{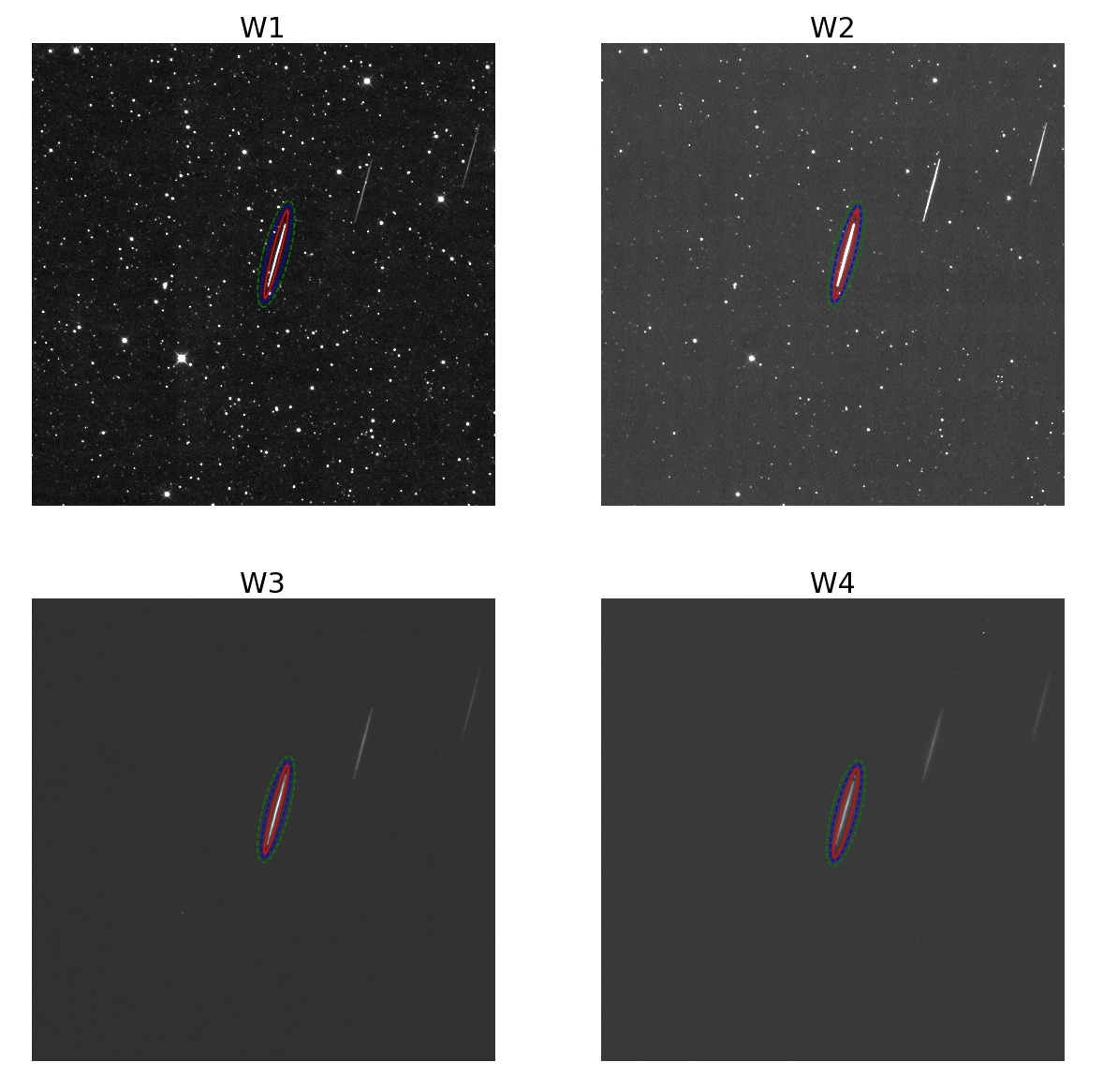}
\caption{An example of an RSO detected in WISE band passes on the 13th February, 2010 at 11:16 UTC. The circled RSO is catalog matched to be Echostar 4, NORAD: 25331, and has a SNR of 181, 430, 1203 and 206. for W1, W2, W3, and W4 respectively for a typical WISE single frame with exposure time of \qty{8.8}{\second} in W3 and W4, and \qty{7.7}{\second} in W1 and W2.}
\label{fig:wise_images}
\end{figure}

To anchor this analysis to current capabilities, we loosely model IR SDA operations on an existing infrared space telescope that has been used in the past to collect intelligence of RSOs, the NASA WISE space telescope.
The images from the WISE survey are all sub-\qty{10}{\second} exposures, however, a dedicated infrared SDA facility could take multiple exposures to integrate on a single GEO target for up to \qtyrange{30}{40}{\minute} per pass on the dark side of the orbit, depending on reasonable Earth limb angle limits. 
Critically, for all analysis in this work we assume a system where all observations are Poisson noise dominated. In order to keep the work generic and applicable to any IR SDA system, we treat uncertainties as Gaussian (assuming a regime of sufficiently high photon counts), and do not assume any specific systematic uncertainties that are associated with IR photometry. An example of RSO imagery acquired by WISE is provided in \cref{fig:wise_images}. This figure presents serendipitously acquired images of Echostar 4 (NORAD catalog ID: 25331), a representative geosynchronous telecommunications satellite.
Given that an RSO has a typical temperature of several hundred Kelvin, its emission peaks at wavelengths longer than \qty{3}{\micro \meter} and is visible in all four WISE filters. 
Hence, wavelengths from $\sim 1-30 \mu m$ for which the WISE filters span, is an appropriate spectral range to focus on (see \cref{fig:wise_bandpasses}).  In the WISE filters, observations with signal-to-noise ratios of SNR $\gtrsim$ \num{1000} are achievable for a typical GEO target with exposure times of \qtyrange{10}{300}{\second}. 
Details of our WISE data analysis and SNR calculations are provided in \ref{sec:apdx:snr}.
\\
\\

\section{Thermal Modeling}\label{sec:thermal}
\begin{figure}
\centering
\includegraphics[scale=1.0]{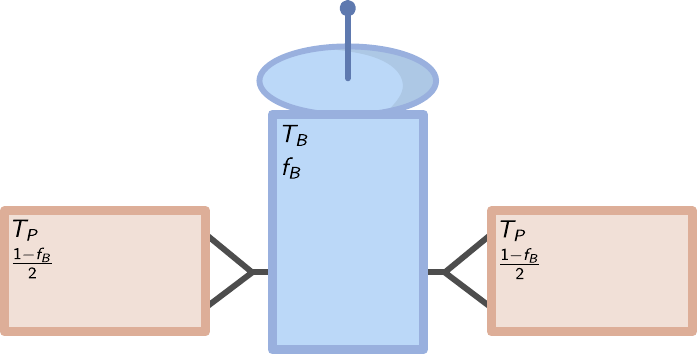}
\caption{
Illustration of a two-component \textqt{box-wing} type satellite with a hot-component of two solar panels at temperature $T_P$ each contributing a proportion of $(1 - f_B)/2$ to the effective area and cool-component of a body at temperature $T_B$ contributing the remaining proportion $f_B$ to the effective area. The IR spectrum of such a satellite is modeled by our two-component spectral model (eqn. \ref{eqn:2bb}) by adding the contribution of the greybody emission at temperature $T_P$ from the solar panels and the greybody emission from the body at temperature $T_B$. When illuminated by the Sun and idle $T_P$ will typically exceed $T_B$ due to their geometry.
}
\label{fig:idea}
\end{figure}

We propose a toy two-component model for a \textqt{box-wing} type satellite consisting of a body at temperature $T_B$ and extended planar solar panels at temperature $T_P$ (\cref{fig:idea}). This is the second most basic model, and simplest meaningful model that displays prospects towards RSO characterization. For example, some one-component models are generally reported to have limited ability of discriminating satellites from space debris and rocket bodies \citep{fitzmaurice_streak_2018, paxson_space_2008}. A two-component model is also the simplest model which can include power-flow between satellite components. We expect that due to the varied geometry of a \textqt{box-wing} type satellite that the presence of a bimodal temperature distribution is a powerful indicator to resolve satellites from other RSO.
\\
\\
Thermal modelling of this basic two-component spacecraft can be used to illustrate feasible solar array and body component temperatures once a specific geometry for the analysis is selected. For this illustrative purpose, we choose the body to be a sphere of radius $R_B$ and the solar panels to be planar with a face area of $A_P$. While the exact temperatures will be different for other convex sphere-like body geometries, to the level of this conceptual study we expect similar temperatures and dynamics to apply. 

Further, we assume that the solar panels are Sun-facing and that the satellite is in Geosynchronous Equatorial Orbit (GEO). Under these conditions, radiation from sources other than the Sun is minimal and not considered in this work. Eclipsed scenarios are also not considered. To evaluate a \textqt{worse-case} of minimum temperature differences solar arrays and chassis, the solar beta angle is taken to be it's solstice maximum of $\psi = \qty{23.4}{\degree}$. Furthermore as a simplifying approximation, we assume that the two model components are not thermally-coupled to each other, neither radiatively nor conductively. However, we introduce constant net power flows into the spacecraft body $P_B$, and solar panels $P_P$ to describe power extraction or use by the components. Under these assumptions, the heat balance equations for the panels and body are given respectively by:
\begin{eqnarray}
  \sigma_B \varepsilon_{P,IR} (2 A_P) T_P^4 =& I_0 \cos(\psi) \varepsilon_{P,vis} A_P &+ \quad P_{P},\nonumber\\
  \underbrace{\sigma_B \varepsilon_{B,IR} (4 \pi R_B^2) T_B^4}_{\text{thermal radiation}} =& \underbrace{I_0 \varepsilon_{B,vis} \pi R_B^2}_{\text{solar irradiance}} &+ \underbrace{P_{B}}_{\text{net power}}. \label{eqn:balance}
\end{eqnarray}

\begin{figure}
\centering
\includegraphics{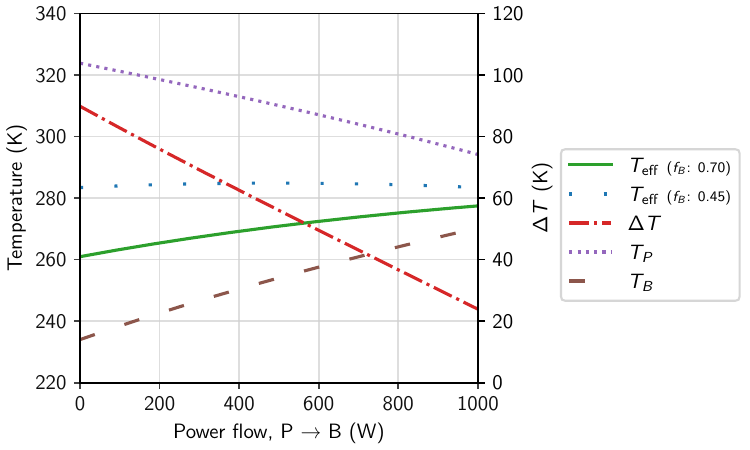}
\caption{Thermal modeling of temperatures for the two-component satellite with body at temperature $T_B$ and solar panels at temperature $T_P$ for varying power flow from the panels to the body ($P_{flow} = - P_P = P_B$) show their different dependence and large temperature difference. The effective greybody temperature of the combined spectra ($T_{\text{eff}}$) from two observation perspectives ($f_B$ values) and temperature difference ($\Delta T = |T_P - T_B|$) are also provided. For $f_B = \num{0.45}$, the effective temperature shows little dependence on power flow.
}
\label{fig:steady-state}
\end{figure}

In these equations $\sigma_B$ is the Stefan-Boltzmann constant, $\varepsilon_{i,j}$ is the emissivity of component $i \in \{P,B \}$ in spectral region $j \in \{vis, IR \}$, and $I_0 = \qty{1360}{\watt \metre^{-2}}$ is the solar irradiance at normal incidence near Earth. 
\\
\\
In this work we model the satellites in both active and idle states. To model a moderately sized satellite we take the body radius to be $R_B = \qty{1}{m}$ and the face area of the solar panels to be $A_P = \pi R_B^2 = \pi \, \unit{\meter^2}$. For simplicity we assume emissivity $\varepsilon_{P,vis} = 0.8,~ \varepsilon_{P, IR} = 0.8, ~\varepsilon_{B,vis} = 0.3,~ $ and $\varepsilon_{B,IR} = 0.6$, typical values for a Kapton surface insulated body and solar cells \citep{gaposchkin_infrared_1995}. Using these numbers, when the satellite is idle ($P_P = P_B = 0$), the temperature of the panels and the body are $T_P = \qty{324}{\kelvin}$ and $T_B = \qty{234}{\kelvin}$, respectively. We highlight that this large temperature difference between the two components of $\qty{90}{\kelvin}$ is solely due to geometry and emissivity; both the sphere and the plane collect the same power from solar irradiance, but the sphere has twice the radiating area.
\\
\\
In the active state, power from the solar panels flows to the satellite body to power its operations. We define $P_{\text{flow}} = P_B = - P_P$ to be the amount of power flowing. The temperature of the two-components at varying power flows is provided in \cref{fig:steady-state}. The temperatures are provided up to $P_{flow} = \qty{1000}{\watt}$ which represents a solar panel efficiency of $\qty{22}{\percent}$ (based on the fraction of solar irradiance incident on the panels). \Cref{fig:steady-state} demonstrates that the effect of power flow is on the same order as that of geometry with the temperature difference $\Delta T = |T_P - T_B|$, reducing to the order of \qty{20}{\kelvin} at the maximum $P_{\text{flow}}$ of $\qty{1000}{\watt}$. 
Also displayed is the effective temperature $T_{\text{eff}} = (1-f_B) T_P + f_B T_B$ of the satellite where the dimensionless factor $f_B \in [0,1]$ accounts for emissivity, area, and view-factors of both components\footnote{$f_B$ is not to be confused with its non-dimensionless counterpart - the radiative exchange factor which is also commonly denoted by $f$.}. 
As justified below, the effective temperature is indicative of the temperature that would best describe this satellite if a single greybody model was fit to its spectrum. Two values of $f_B$ are shown, each representing different observation perspectives.
For $f_B = 0.7$, a strong variation with power flow is seen in both $T_{\text{eff}}$ and $\Delta T$, however there exist observation perspectives where variation in $T_{\text{eff}}$ with power flow is largely canceled by the opposite increase of $T_B$ and decrease of $T_P$. This is shown with $f_B = \num{0.45}$ where the adversarial value of $f_B$ reduces the variation in $T_{\text{eff}}$ to only within a few Kelvin. Unlike $T_{\text{eff}}$, $\Delta T$ exhibits the same dependence on power flow regardless of the perspective the RSO is observed from. Thus, the ability to resolve $\Delta T$ provides a robust insight for RSO characterization and monitoring.

This basic thermal modeling provides insight into the parameter range of interest for RSOs in GEO. For RSOs with planar and sphere-like components the effective temperature will be around $\approx \qty{300}{\kelvin}$ and the temperature difference, $\Delta T$, will be on the order of 10s of Kelvin. 

\section{Spectral Modeling}\label{sec:theory}
\subsection{A two-component satellite spectral model}

Informed by the assumptions and framework presented in the previous section, we propose a corresponding two-component spectral model for the satellite's spectra in the thermal IR wavelength range where the radiation is dominated by thermal emission.
In our two-component model, the spectra is a linear combination of a hot temperature ($T_H$) and cold temperature ($T_C$) greybody spectra. 
These represent the emissions from the two components of the satellite, e.g.: the satellite body, and extended solar panels. Recall for the thermal modeling in the previous section the analysis required we chose some specific satellite geometry (e.g. a sphere for the chassis and planar solar panels); in contrast, our spectral model has no inherent regard for the geometry of the emitting spacecraft components, is only directly dependent on the component's temperatures and their effective areas to the observer. Furthermore, in the thermal modelling we considered the temperatures $T_P$ and $T_B$ of the solar panel and body; for the spectral model, the component temperatures are first inferred from data without a direct association to the emitting RSO component. In our spectral model we therefore work with the cooler $T_C$ and warmer $T_H$ greybody temperatures directly rather than the temperatures of specific components (i.e. $T_P$ and $T_B$).
Finally, to ease theoretic analysis and emphasize properties of the two-component spectrum more independent of RSO-component size, observation range, and RSO attitude such as temperatures, we further consider a non-dimensionalised version of the spectral intensity for this model given by:
\begin{eqnarray}
I(\lambda, \alpha, f_C, T_{\text{eff}}, \Delta T) = \alpha \Bigl(f_C B\bigl(\lambda, T_C(f_C, T_{\text{eff}}, \Delta T)\bigr) + (1 - f_C) B\bigl(\lambda,T_H(f_C, T_{\text{eff}}, \Delta T)\bigr) \Bigr),\label{eqn:2bb}
\end{eqnarray}
where
\begin{eqnarray}
T_C(f_C, T_{\text{eff}}, \Delta T) =& T_{\text{eff}} - (1 - f_C)\Delta T\\
T_H(f_C, T_{\text{eff}}, \Delta T) =& T_{\text{eff}} + f_C \Delta T\\
B(\lambda,T) =& \lambda^{-3} \left(\exp\left(\frac{\hbar c}{2 \pi \lambda T} \right) - 1 \right)^{-1}
\end{eqnarray}
is a non-dimensional Planck distribution. 
Here, $\lambda$ is the wavelength, $\alpha$ is the effective area, $f_C$ is the proportion the cold greybody contributes to the effective area of the satellite, $T_{\text{eff}}$ is the effective temperature of the satellite, and $\Delta T$ is the temperature difference between the hot and cold greybody components with temperatures $T_C = T_{\text{eff}} - (1 - f_C)\Delta T$ and $T_H = T_{\text{eff}} + f_C \Delta T$ respectively. 

In terms of physical parameters of the emissivity of the hot and cold components $\varepsilon_H$, $\varepsilon_C$, and their projected areas from the observer $A_H, A_C$ the model parameters $\alpha$ and $f_C$ are given by:
\begin{eqnarray}
  \alpha =& k(\varepsilon_C A_C + \varepsilon_H A_H),\\
  f_C =& k\varepsilon_C A_C/\alpha = 1 - k\varepsilon_H A_H/\alpha,
\end{eqnarray}
where $k$ is a normalization coefficient which makes $\alpha$ non-dimensional. Moreover we note that in terms of physical RSO parameters, $\Delta T$ depends only on the hot and cold component temperatures ($T_C$ and $T_H$) and not on their projected areas and emissivities. $\Delta T$ is therefore a promising signature for RSO characterization as it is not directly dependent on RSO geometry or surface materials and is invariant to changes in the observers perspective, RSO attitude, and range.

This two-component model does not include any nuisance terms for spectral contributions such as reflected Sunlight or Earthshine. Given the dimming factor due to the limited angular extent of the Sun, its spectral contribution in the IR bands $\left(\frac{R_{\odot}}{\mathrm{AU}}\right)^2  B(u, T_{\odot}) $ is generally larger yet comparable to thermal emissions in the WISE W1 and WISE W2 bands and smaller than thermal in WISE W3 and W4. For a blackbody at \qty{300}{\kelvin} the ratio of reflected spectral irradiance to blackbody spectral irradiance is \numlist{27; 1.0; 5E-3; 1E-3} at W1 - W4 respectively. Thus, in a more detailed model where the surface properties of the satellite and different illumination angles are considered, it is important to include the reflected component, particularly for filters centered at shorter wavelengths such as W1 or W2. Correction for these effects are left to future work dedicated to WISE observation analysis.

\subsection{Comparison to the one-component spectral model}

In the limit as the temperature difference between components of the satellite $\Delta T$ approaches 0 the two-component model becomes equivalent to the one-component greybody model:
\begin{eqnarray}
  I_1(\lambda, \alpha, T) = \alpha B(\lambda, T_{\text{eff}}), 
\end{eqnarray}
and as expected, in this limit the value of $f_C$ also becomes degenerate. 
Given we expect to operate in the small ($\lessapprox \qty{100}{\kelvin}$) $\Delta T$ regime,
we can view the parameter pair $(\alpha, T_{\text{eff}})$ as corresponding to the properties of an effective single greybody model and the parameter pair $(f_C, \Delta T)$ as corresponding to the properties of the correction resolving the spectra into two greybody components. Applying Taylor's approximation in the greybody temperature, it can be shown that the mixture model is equal to first order in the temperature difference $\Delta T$ to a single greybody spectral model with effective temperature, $f_C T_C + (1 - f_C) T_H$.
This provides a motivation for considering the $T_{\text{eff}} = f_C T_C + (1 - f_C) T_H$ parameter of our model to be the temperature of an effective single greybody. Furthermore, it tells us that the one- and two-component greybody models will differ only at most quadratically in $\Delta T$. 
As a consequence, it will be seen that the parameter pair which resolve the two-greybody, $(\Delta T, f_C)$, will be much more sensitive to measured IR-band values than the parameters which determine the effective greybody, $(\alpha, T_{\text{eff}})$.

\section{Numerical Analysis}\label{sec:numerics}

\subsection{3D color diagram}
\begin{figure}
\centering
\includegraphics[scale=1.0]{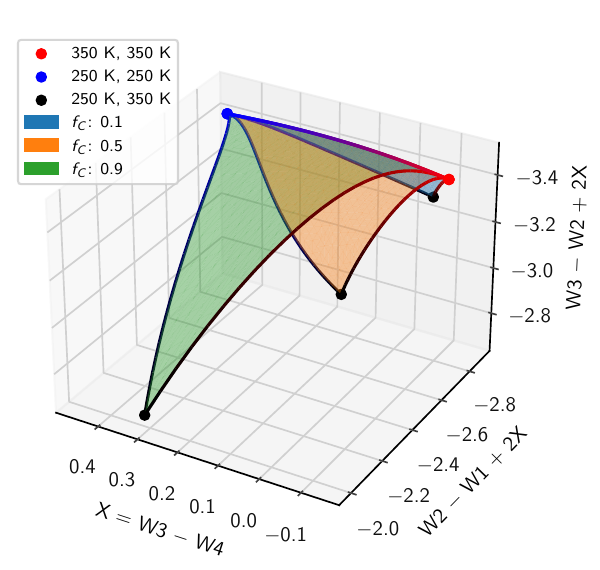}
\caption{A 3D color diagram for the two-component greybody spectral model (eqn. \ref{eqn:2bb}) over parameter range ($f_C \in [0,1], T_C \in [\qty{250}{\kelvin}, \qty{350}{\kelvin}], T_H \in [T_C, \qty{350}{\kelvin}]$). Each axis gives the magnitude of the respective WISE band-pairs with the $y$-axis and $z$-axis sheared according to $y \leftarrow y + 2x$, and $z \leftarrow z + 4 x$ respectively for visualization purposes. 
Constant $f_C$ surfaces are shown in blue, orange, and green and boundary points for exact temperature of the hot and cold components are given in form: $(T_C, T_H)$ in the legend.
The injectivity of the volume into this multidimensional color space indicates the possibility to solve for the model parameters from WISE band measurements.
}
\label{fig:color3}
\end{figure}

For a typical (physically motivated) parameter range ($f_C \in [0,1], T_C \in [\qty{250}{\kelvin}, \qty{350}{\kelvin}], T_H \in [T_C, \qty{350}{\kelvin}]$ ), the monochromatic flux at the central wavelength of the four WISE filters (W1 - W4) are sufficient to determine the four model parameters. Because the color-index in any two bands is independent of the effective area $\alpha$, it is possible to visualize how the parameter space maps into a 3-dimensional space of color-indices as done in \cref{fig:color3}. This diagram extends the idea of using color-color diagrams for space domain awareness explored in \citet{reyes_analysis_2023, fitzmaurice_streak_2018, fitzmaurice_detection_2021} into 3-band pairs - an idea also explored in \citet{kielkopf_remote_2022}. We term this type of diagram a 3D color diagram.

In \cref{fig:color3} three constant $f_C$ surfaces of the parameter space have been drawn. We recall that small values of $f_C$ correspond to a low relative effective area for the cold component at temperature $T_C$. As expected, all the surfaces intersect when $\Delta T = \qty{0}{\kelvin}$ at the blue-red colored boundary.  $\Delta T$ increases perpendicular to this boundary and, agreeing with theory, the $f_C$ surfaces separate at most quadratically in $\Delta T$ from this boundary.

Without the axis detrending used in \cref{fig:color3}, the parameter space is contained within a tight neighborhood along the curve of varying effective temperature $T_{\text{eff}}$ for the spectra. That is, changes in $T_{\text{eff}}$ cause much larger movements in parameter space compared to $f_C$ or $\Delta T$.
After detrending, the physical parameter space occupies a volume of 3D color space approximately \num{0.5} magnitude in side-length and with the relative magnitude of down to $\num{-3.4}$ needed to cover the low-temperature ($T_{\text{eff}} = \qty{250}{\kelvin}, \Delta T = \qty{0}{\kelvin}$) region of the parameter space. This gives an order-of-magnitude indication of the precision of the photometry needed to locate a measurement in parameter space, and hence to infer the parameter values.

\subsection{Parameter estimation uncertainty}\label{sec:numerics:uncert}

Key to practical use of any spectral model in SDA is the ability to infer its parameters from observations robustly and without significant variance or systematic biases. While the analysis of systematic errors introduced in photometric analysis is beyond the scope of this work (for example, errors in zeropoint calibration or confusion limits from nearby stars), in order to  develop an understanding of the uncertainty associated with parameter inference from measurement and the main factors affecting this for our two-component model, we now study the sampling distribution of the model parameters. This sampling distribution gives the probability of obtaining given model parameters from repetitions of the same observation experiment.

In this work, we use WISE as a reference example of the central wavelength and expected SNR for an infrared space telescope. However, we stress that we do not consider any WISE specific noise properties or systematic uncertainties due to the complexities of infrared photometry beyond assuming typical SNR ratios that we demonstrate are feasible based on actual WISE observations (cf. \cref{fig:wise_images}). A fullons or reducing the number of fields/variables or restricting them to a particular symmetric for description of our Monte Carlo method is given in \ref{sec:apdx:methods:uncert}.
In the presentation of our results, we adopt the notation of adding the $\cdot_0$ subscript to indicate true parameter values and adding the $\hat{\cdot}$ diacritic to indicate estimated parameter values.
One challenge of fitting this two-component model to observation is that the data are well-described by a one-component model in the limit that $\Delta T$ approaches $0$. In this case, one component of the two-component model will describe the temperature and effective area of the RSO and be physically meaningful while the second component will fit the noise-dominated residuals and thus have a temperature and effective area which do not give physical information about the RSOs state. Criterion for determining if obtained fit parameters are physically meaningful are presented and discussed in \ref{sec:apdx:methods:uncert}.

\begin{figure}
\centering
\includegraphics[scale=1.0]{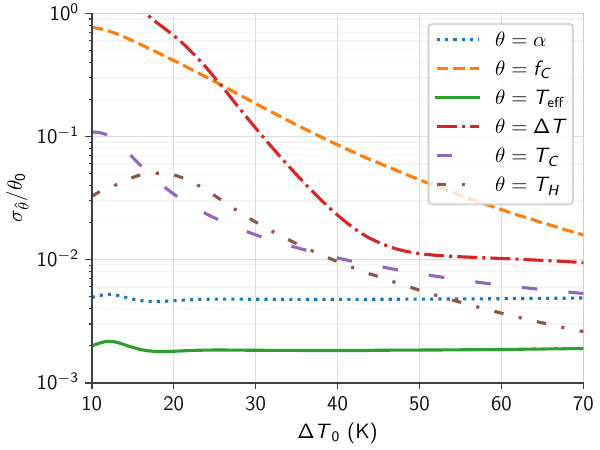}
\caption{
  Relative uncertainties for model parameters of $\alpha_0 = 1, {f_C}_0 = 0.7, {T_{\text{eff}}}_0 = \qty{300}{\kelvin}$, and varying ${\Delta T}_0$ calculated using WISE filters with $\sigma_I / I = \num{E-3}$ relative uncertainty (SNR \num{1000}) in each observation band.
  For the model parameters which resolve the two greybody components ($f_C$, $\Delta T$) the uncertainty decreases rapidly with increasing ${\Delta T}_0$. The plot shows that for ${\Delta T}_0\gtrsim \qty{30}{\kelvin}$ the uncertainties are sufficiently small to enable inference on RSO characterization.  
}
\label{fig:freq}
\end{figure}

\Cref{fig:freq} displays the parameter uncertainty for varying ${\Delta T}_0$ from measurements of the spectra in each of the IR WISE bands. Firstly we note that during the estimation of these sampling distributions, a physical fit of the two-component model is always achieved for ${\Delta T}_0 > \qty{30}{\kelvin}$, where the bolometric flux of each component is non-negligible. After this point the number of nonphysical fits increases to $3\%$ of all simulated observations by ${\Delta T}_0 = \qty{20}{\kelvin}$ and $\qty{50}{\percent}$ by ${\Delta T}_0 = \qty{10}{\kelvin}$. This high proportion of nonphysical fits at very low ${\Delta T}_0$ indicate that fitting the two-component model with four bands and the given SNR is insufficient, and further suggests that the one-component model may be becoming more meaningful here. Detailed results discussing model preference are provided below in \Cref{sec:numerics:compare}.
Again, this behavior is expected; as ${\Delta T}_0 \rightarrow 0$ the spectrum approaches that of a single greybody which only requires two parameters for specification rather than the four fit here. Hence, in this limit, the parameters of the two-component model are less informed by the data.

Of the four model parameters, $\alpha$ and $T_{\text{eff}}$ can be determined with higher precision than $f_C$ and $\Delta T$. 
This is especially true at lower ${\Delta T}_0$; when ${\Delta T}_0 = \qty{30}{\kelvin}$, $\hat{\alpha}$ and $\hat{T_{\text{eff}}}$ have uncertainties of $\num{4.8E-3}$ and $\qty{0.55}{\kelvin}$ respectively, whereas $\hat{f_C}$ and $\hat{\Delta T}$ have uncertainties of $\num{0.13}$ and $\qty{3.5}{\kelvin}$ respectively. 
This agrees with the theoretical analysis of the two-component model and supports the fact that while the effective greybody properties can be easily measured, resolving the two greybodies, while possible, is more challenging.

\begin{figure}
\centering
\includegraphics[width=\textwidth]{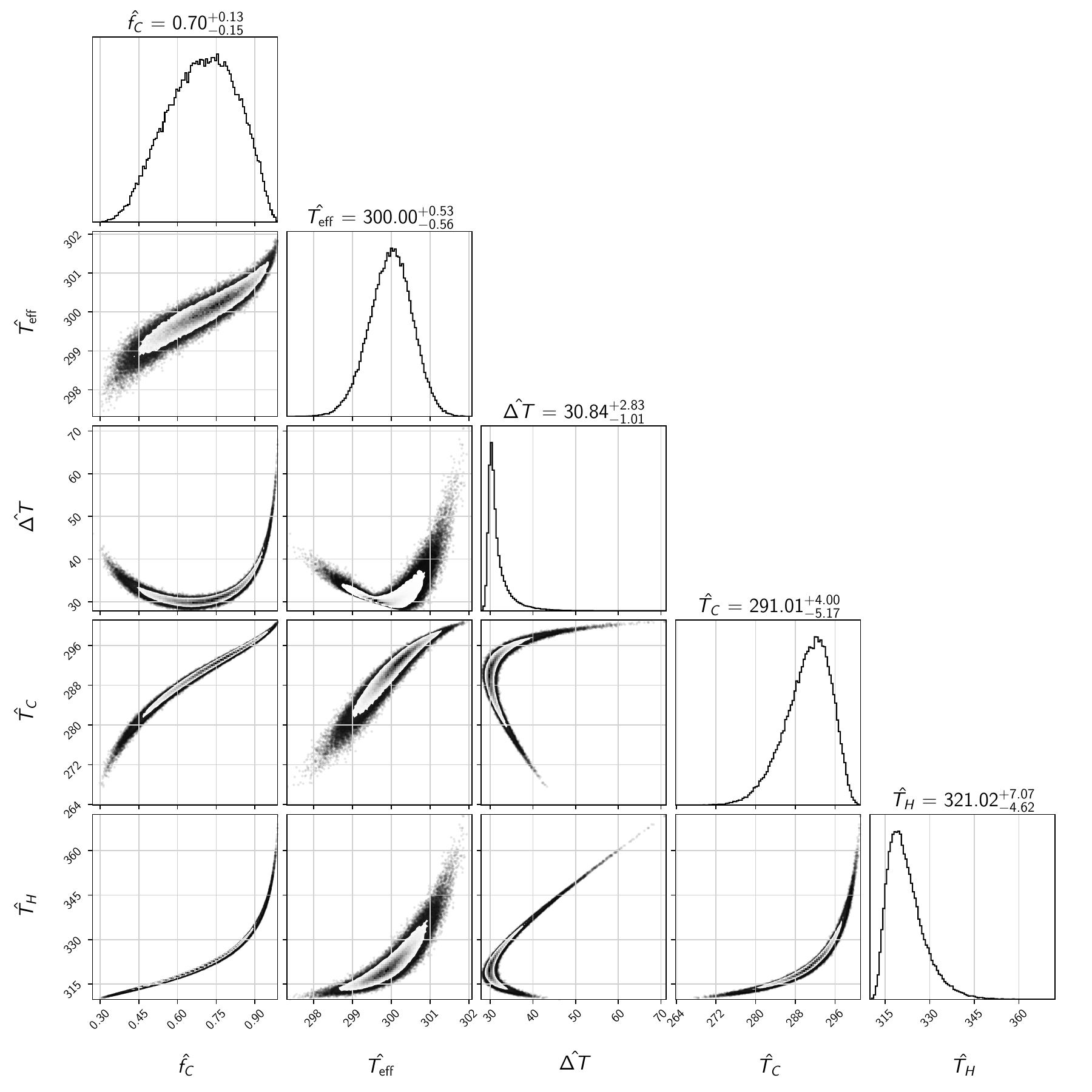}
\caption{Corner plot for parameter uncertainty with $\alpha_0 = 1, {f_C}_0 = 0.7, {T_{\text{eff}}}_0 = \qty{300}{\kelvin}$, and ${\Delta T}_0 = \qty{30}{\kelvin}$ calculated using WISE central wavelengths, with $\sigma_I / I = \num{E-3}$ relative uncertainty (SNR 1000) at each central wavelength. This joint sampling distribution highlights the strong dependence between parameters and thus the benefit to parameter uncertainty that could be obtained from independent constraints such as prior knowledge of the geometry of the RSO.
}
\label{fig:corner}
\end{figure}

The estimated component greybody temperatures $\hat{T_C}, \hat{T_H}$ have uncertainties of $\qty{5}{\kelvin}$ and $\qty{6}{\kelvin}$ respectively which is larger than the uncertainty in $\hat{\Delta T}$, and also decreases with increasing ${\Delta T}_0$. The uncertainties in $\hat{T_C}$ and $\hat{T_H}$ are about an order of magnitude larger than $\hat{T_{\text{eff}}}$ at ${\Delta T}_0 = \qty{30}{\kelvin}$, however they approach the uncertainty in $\hat{T_{\text{eff}}}$ as ${\Delta T}_0$ increases through to $\qty{70}{\kelvin}$.
\\
\\
To gain a better understanding of the dependence between the model parameters, the joint sampling distribution at ${\Delta T}_0 = \qty{30}{\kelvin}$ is presented in \cref{fig:corner}. While \cref{fig:steady-state} demonstrates ${\Delta T}_0$ values up to $\qty{90}{\kelvin}$, as the uncertainties were generally observed to increase with decreasing ${\Delta T}_0$, the value of ${\Delta T}_0 = \qty{30}{\kelvin}$ is chosen to inform on a more challenging case.
The sampling distribution in \cref{fig:corner} was obtained similarly to the sampling distribution used to generate \cref{fig:freq} except with a greater number of experiments of \num{E5}.

The obtained distributions in \cref{fig:corner} are visibly non-Gaussian. Moreover, a positive correlation is observed between the parameters of $\hat{T_{\text{eff}}}$ and $\hat{f_C}$ and a strong but non-monotonic relationship is observed between $\hat{\Delta T}$ and $\hat{f_C}$, and $\hat{\Delta T}$ and $\hat{T_{\text{eff}}}$. 
That is the relationship between these variable pairs is neither only increasing or decreasing and thus certain values of $\hat{\Delta T}$ there exist disjoint regions of possible values of $\hat{f_C}$ and $\hat{T_{\text{eff}}}$.
Furthermore, we see that $\hat{\Delta T}$ has a strong positive skew and the majority of estimates for $\hat{\Delta T}$ are an overestimate. The number of degrees of freedom in our spectral model reduces from four to two as ${\Delta T}_0$ approaches zero, which may contribute to the observed tendency to overestimate ${\Delta T}_0$.

This non-convexity between $\hat{\Delta T}$ and $\hat{f_C}$ implies that a prior on ${\Delta T}_0$ (centered around the true value) is in general not sufficient to further constrain $\hat{f_C}$ without degeneracy.
At lower ${\Delta T}_0$, the observed dependence explains a large part of the uncertainty. At ${\Delta T}_0 = \qty{30}{\kelvin}$ an additional independent constraint on $\hat{f_C}$ (such as from the geometry of a known RSO) with uncertainty $\qty{10}{\percent}$ reduces the uncertainty in $\hat{T_{\text{eff}}}$ and $\hat{\Delta T}$ by \qty{47}{\percent} and \qty{68}{\percent} respectively.

\begin{figure}
\centering
\includegraphics[scale=1.0]{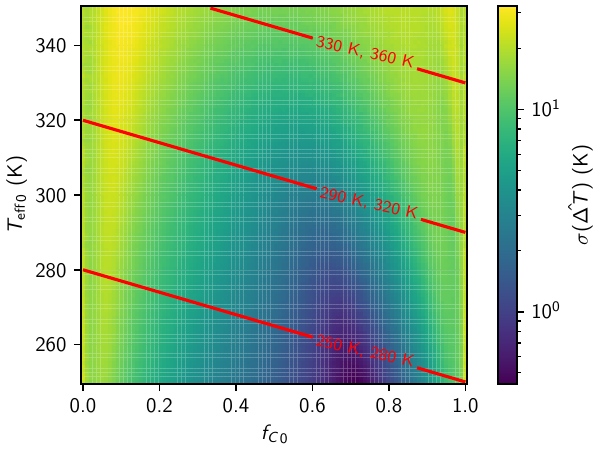}
\caption{
  Parameter uncertainty in $\hat{\Delta T}$ for model parameters of $\alpha_0 = 1$ and ${\Delta T}_0 = \qty{30}{\kelvin}$, and varying ${f_C}_0$, and ${T_{\text{eff}}}_0$ calculated using WISE central wavelengths with $\sigma_I / I = \num{E-3}$ relative uncertainty (SNR \num{1000}) in each observation band.
  Temperature contours of the hot and cold greybodies are shown in red and are labeled in the format ${T_C}_0, {T_H}_0$. The plot shows that the uncertainty in $\hat{\Delta T}$ varies by up to two orders of magnitude with varying ${f_C}_0$ and implies that the minimum uncertainty is obtained near ${f_C}_0 \geq 0.5$ and rapidly grows at ${f_C}_0 \rightarrow 0$ or ${f_C}_0 \rightarrow 1$.
}
\label{fig:DT_var}
\end{figure}

Finally, we see that the sampling distribution between $\hat{f_C}$ and $\hat{T_H}$ and to a lesser extent $\hat{T_C}$ conform to tight distributions. These distributions are tighter than the distributions between $\hat{T_C}$ and $\hat{T_H}$ and either of $\hat{T_{\text{eff}}}$ or $\hat{\Delta T}$. From this, we can also conclude that independent constrains on $\hat{f_C}$ would also be best to constrain both $\hat{T_C}$ and $\hat{T_H}$.
\\
\\
Of the spectral parameters, $\Delta T$ is highly informative on RSO behavior but more challenging to infer. \Cref{fig:DT_var} displays how the parameter uncertainty in $\hat{\Delta T}$ varies with the true effective temperature ${T_{\text{eff}}}_0$ and cold-blackbody fraction ${f_C}_0$\footnote{Due to the large number of parameter uncertainty tasks needed to generate this figure, \num{10000} experiments were used for each ${T_{\text{eff}}}_0, {f_C}_0$ point.}. At each ${T_{\text{eff}}}_0$ in the plot, a minima in the uncertainty is obtained at between $0.5 \leq {f_C}_0 \leq 0.8$, when the cold greybody has a greater effective area than the hot greybody.
As ${f_C}_0$ approaches $0$ or $1$, the uncertainty in $\Delta T$ greatly increases before leveling and slightly decreasing from the peak. The increase can be understood from the fact that very little flux from one of the greybodies is present in this regime which makes fitting a spectra to it challenging. The leveling and decrease occurs as beyond this point, the large number of nonphysical fits (where the flux from one of the components is less than \qty{1}{\percent} of the total) prevents $\hat{\Delta T}$ from growing without bound; we note this while also noting that a two-greybody spectral fit is no longer a meaningful model in this regime.
\\
\\

\subsection{Model preference}\label{sec:numerics:compare}

As outlined in \cref{sec:theory}, our two-component satellite spectral model approaches the one-component model in the limit that ${\Delta T}_0$ approaches zero, or ${f_C}_0$ approaches zero or one. It is therefore natural to ask for a given set of sampled wavelengths and uncertainties for what parameter domain (${\Delta T}_0, {f_C}_0$) is it meaningful to decompose the spectra into a two-component model. That is, when is the two-component model a better fit to the data than the one-component model?

\begin{figure}
\centering
\includegraphics[scale=1.0]{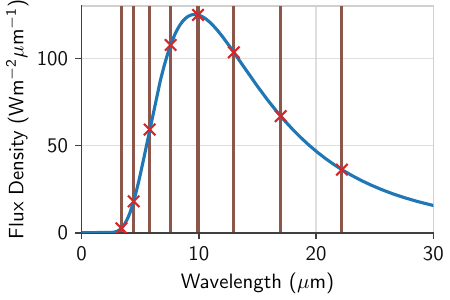}
\caption{
  The eight logarithmically spaced filters for model preference analysis (brown vertical lines) sampling an indicative two-component spectrum with ${f_C}_0 = \num{0.7}$, ${T_{\text{eff}}}_0 = \qty{300}{\kelvin}$, and ${\Delta T}_0 = \qty{10}{\kelvin}$.
}
\label{fig:logbands}
\end{figure}

Details of the used model preference test are provided in \ref{sec:apdx:methods:preference}, however note that eight filters logarithmically spaced between WISE W1 and W4 (\cref{fig:logbands}) were used.
Eight filters, as opposed to four, were used have a non-zero number of degrees of freedom in the fits required for meaningful model preference comparison.
Logarithmic, as opposed to linearly, spaced samples were chosen as logarithmic spacing displayed a many fold reduction in parameter uncertainty. We postulate that one reason for this may be that logarithmically spaced samples more evenly sample both tails of the spectrum.

\begin{figure}
\centering
\includegraphics[scale=1.0]{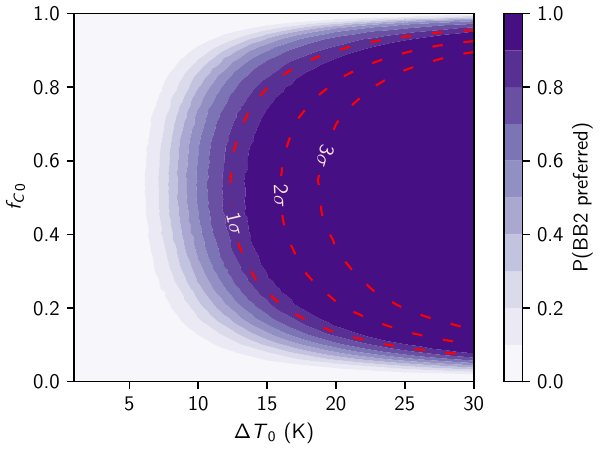}
\caption{
  Probability that an observation prefers the two-component model over the one-component model for eight measurement filters with relative uncertainty $\sigma_I/I = \num{e-3}$ logarithmically spaced between WISE W1 and W4 (\cref{fig:logbands}) at $\alpha_0 = 1$ and ${T_{\text{eff}}}_0 = \qty{300}{\kelvin}$. Given the expected temperature difference $({\Delta T}_0)$ range for RSO extends to almost \qty{100}{\kelvin} (\cref{fig:steady-state}) this plot demonstrates that an RSO observation can robustly distinguish between the spectrum of a one-component and two-component RSO for the majority of expected two-component spectra.
}
\label{fig:preference}
\end{figure}

\Cref{fig:preference} presents the probability that an observation prefers the two-component model. 
The two-component model is preferred for the majority of the parameter space of interest as given by our thermal modeling. This implies that for an RSO observation in eight bands at an SNR of 1000 considering Poisson noise dominated observations, it is generally meaningful to decompose the spectra into our two-component model.
Specifically, the two-component model is preferred when ${\Delta T}_0 \gtrapprox \qty{15}{\kelvin}$ and ${f_C}_0$ is not close to $0$ or $1$, and demonstrates robust capacity to resolve a two-component RSO.
Reassuringly, the one-component model begins to be preferred at a similar boundary to when the values of ${\Delta T}_0$ or ${f_C}_0$ which make the two-component model degenerate with the one-component model are within the uncertainty of $\hat{\Delta T}$ and $\hat{f_C}$.

\section{Discussion}\label{sec:discussion}

Extending previous works \citep{paxson_space_2008, fitzmaurice_streak_2018, wright_infrared_2023}, our RSO spectral model resolves the IR spectra into two greybody components. In the context of RSO characterization, the new parameter of the greybody temperature difference, $\Delta T$, provides unique, perspective invariant, insight into the RSO's operations. 
A consequence of the demonstrated ability to resolve ${\Delta T}_0$ in a two-component model is that changes in power-flow between the components can be sensitively monitored. Combining the thermal modelling of the toy satellite with the parameter uncertainty results, we estimate that measurement in a few photometric band at an SNR of 1000 and of an RSO with component temperature difference (${\Delta T}_0$) greater than $\qty{30}{\kelvin}$, changes in the power-flow of between $\qtyrange{5}{10}{\percent}$ can readily be obtained. Such a resolution on the power flow would be sufficient for basic monitoring of the operational status of an RSO. We remark that while variability is high (both on orbit and between spacecraft), a $\Delta T$ of $\qty{30}{\kelvin}$ is fairly representative for many satellites; such or larger a $\Delta T$ is seen between the chassis and deployable solar panels of cubesats \citep{blanchete_thermal_2024, mason_cubesat_2018} and in larger satellites such as between the baffle and chassis of the $\approx \qty{10}{\meter}$ Hubble space telescope \citep{nino_hst_2008}.
\\
\\
Observing changes in the power-flow is dependent on ${\Delta T}_0$, yet its estimation becomes increasing difficult as ${\Delta T}_0$ decreases especially below $\qty{20}{\kelvin}$ where at the presented SNR the relative uncertainty in $\hat{\Delta T}$ approaches unity.
We find that at each ${T_{\text{eff}}}_0$ a minima in uncertainty in $\hat{\Delta T}$ is obtained with a cold-greybody fraction greater than \qty{50}{\percent}. For a box-wing type RSO with Sun-pointing solar panels, this may imply that observations are best made near a solar phase angle of $\qty{90}{\deg}$.
\\
\\
The WISE IR wavelength range explored here was chosen as it represented a feasible and appropriate measurement wavelength range for temperature determination. While further extending to longer wavelengths aids in reducing parameter uncertainty -- adding a single extra filter at \qty{30}{\micro \meter}, above WISE W1 to W4, decreases the uncertainty in $\Delta T$ and $f_C$ found from this analysis by \qty{27}{\percent} and \qty{14}{\percent} respectively.

Meanwhile, measurements at wavelengths shorter than W1 are likely of limited benefit as they contain very little greybody spectral radiance. For a greybody at the hotter side of expected RSO temperatures at ${T_{\text{eff}}}_0 = \qty{400}{\kelvin}$ (hence radiating more at shorter wavelengths), the spectral radiance at $\qty{2.7}{\micro \meter}$, immediately below the WISE W1 filter, is only $\qty{13}{\percent}$ that at W1 and $\qty{0.5}{\percent}$ the peak spectral radiance at $\qty{7.2}{\micro \meter}$.
Within the WISE IR wavelength range, our results indicate that the WISE filters are well chosen for precise parameter inference of our two-component model, and that logarithmic sampling of the spectral radiance is generally superior to linear sampling. Both the WISE and the logarithmically spaced filters have preferential sampling of the spectrum at shorter wavelengths, suggesting that spectral measurements at these lower wavelengths are very informative about the parameters.
\\
\\
Our results also highlight the benefit of decreasing parameter uncertainty which is derived from additional independent constraints on the parameters. In particular, priors on ${f_C}_0$ greatly aid in reducing the uncertainty in $\hat{\Delta T}$.
In practice, a constraint on $\hat{f_C}$ could be derived from prior knowledge about the geometry and orientation of a given RSO. Similarly, as ${\Delta T}_0$ is perspective invariant, simultaneous observations from a constellation of orbiting IR SDA telescopes could be combined to constrain this parameter. 
Additional information, either a greater number of filters or external constraints, could also be used to achieve the same parameter uncertainty at a lower SNR requirement or with fainter targets.

\subsection{Prospects for inference of high-level RSO structural information}

Conversely to using prior knowledge about the geometry of an RSO for parameter estimation, unresolved IR spectral characterization of RSOs also has the potential to unveil spatial and high-level structural information about the RSO. We have demonstrated that it is in principle possible to resolve the effective area fraction, ${f_C}_0$, of the two components of the RSO purely from spectral measurement. 

A traditional spatially-resolved optical measurement of the RSO's area ratio using a sub-meter diameter aperture space telescope as investigated in this study would have a diffraction-limited resolution on the order of \qty{1}{\meter} for RSO in LEO (at distance $\approx \qty{1000}{\kilo \meter}$) and \qty{10}{\meter} in GEO (at distance $\approx \qty{40000}{\kilo \meter}$). 
These resolutions are similar to the diameter of RSOs of interest in each of these orbits and therefore, a similar space telescope  targeted at optical wavelengths would be unable to perform spatially resolved measurements.

In the most primitive example, a spherical RSO's spectrum would be invariant to changes in perspective (assuming uniform surface properties), whereas the magnitude of a planar part of the RSO's spectrum would vary with the cosine of the angle between the surface normal and the observer. In a composite RSO consisting of a sphere and a plane, the angular dependence of the spectra from each could be resolved with our two-component greybody model in the case that the two components do not occlude each other.
Observing the total thermal IR flux of satellites in GEO by \citet{skinner_ir_2007} has revealed variations with phase angle between individual satellites, which corroborates this idea.
\\
\\
In general, the inverse problem of inferring geometry from the angular dependence of spectra is a challenging one and likely an ill-posed problem once complex geometries, occlusion, and surfaces' angular dependence of emissivity are considered.
Possible inroads may be made by matching the observed angular dependence to a precompiled library from simulations of typical RSOs rather than attempting a direct inference of the geometry. A full consideration of this problem and its possible solutions are beyond the scope of this work. Furthermore, as the spectra in the optical region is dominated by reflections from the Sun, the angular dependence of the RSOs spectra in the optical region depends also on the position of the Sun, and the surface optical properties of the RSO. This differs from the infrared \citep{wright_infrared_2023} and hence complimentary unresolved optical observations of the RSO may provide additional information for geometric inference.

\subsection{Extension to additional components and prospects for maneuver detection}

While further extension of this model to three or more near-temperature greybodies could reveal more detailed operational and geometric information about the RSO, such an extension is only useful if the temperatures of every greybody can be resolved. This is a challenge in two respects: a) as the range of temperatures remains approximately constant the greybody temperatures become more closely spaced, and b) given the significant extra difficulty in resolving the temperature difference for two greybodies ($\Delta T$) over a single greybody ($T_{\text{eff}}$) we expect the scaling to additional greybodies to be challenging. Both these items suggest that an even greater measurement SNR would be required to resolve more than two greybodies from spectral data alone.
\\
\\
Another potential area for future work would be to explore spectroscopic measurements. Tracking of an RSO could provide the opportunity perform measurements using an instrument similar to that of JWST's mid infrared instrument low resolution spectrometer \citep{kendrew_mid-infrared_2015}, or the Spitzer space telescope's infrared spectrograph \citep{houck_infrared_2004}) with integration times exceeding \qty{300}{\second}. High spectral resolution provided by instruments like these may improve model estimates and offer an interesting additional element to explore for future SDA instrumentation design. Intriguingly, studies of galaxy spectra from the SDSS survey that have been analysed with models at \qty{3}{\angstrom} spectral resolution have shown the ability to reconstruct individual star formation histories of galaxies across cosmic time, i.e. the contributions to the spectrum from many black bodies with different characteristic temperatures are reconstructed successfully \citep{Panter2007}. Qualitatively translating those results to SDA, one may expect to be able to constraint multiple components of a man-made RSO, possibly sufficient to build a low-to-medium fidelity thermal model. Future analysis is clearly needed for further quantitative assessment.

Last, our results demonstrate that resolving individual greybodies from the RSO's spectra becomes easier with increases to the temperature difference. 
Therefore, inferring the presence of a component heated a few hundred Kelvin above the ambient RSO temperature from unresolved spectral observations would be less challenging than the task undertaken here of resolving two components with a much smaller temperature difference.
One main forecast challenge is that reflected solar radiation becomes greatly dominant at wavelengths below WISE W1.
Due to the power required to maintain a component of such high temperature, its presence would be a strong indicator of activity on the RSO. It may, for example, be an actively firing electric thruster \citep{myers_hall_2016}, motivating further study to investigate the feasibility into this capability.

\subsection{Limitations}
The main limitations of our two-component spectral model presented here are: a) variations in $f_C$ with wavelength and material, b) only two greybody components are modeled, c) thermal gradients within components have both been neglected and d) we do not consider systematic uncertainties present in real observations by infrared space telescopes. Our model only captures two spectral components of an RSO, yet a detailed thermal model for an RSO could contain upwards of hundreds of elements, each with their own temperature and spectral radiation. Of these, major components include radiators, resolving illuminated vs shadowed chassis surfaces, high-power antenna and external instrumentation, and multi-layer insulation. Modelling and inference of the temperatures of these additional spacecraft components would supplement the information satellite operators receive via telemetry, though as above, even in a theoretically ideal setting, resolution at this detail would likely go beyond what is feasible given the instrumentation capabilities. 
\\
\\
The two-component nature of our model approximates thermal gradients across these components with a single temperature. For gradients within a few Kelvin, the first-order equality of a mixture of greybody spectra and the found resolution limit for distinct greybodies of near $\qty{15}{\kelvin}$ for the measurement scheme used in this paper both suggest that the true spectra generated by the integral of greybodies across the thermal gradient are well approximated by a single greybody spectra. Thermal gradients larger than a few Kelvin are likely to play an important role when fitting our model. Given heat flow is driven by the gradient of temperature, these thermal gradients will approximately grow with the component's radius with characteristic scale $\frac{\mathrm{d} T}{\mathrm{d} x} = \Phi/\kappa$, where $\Phi$ is the heat flux and $\kappa$ is the component's thermal conductivity. Using a value of $\kappa = \qty{240}{\watt \meter^{-1} \kelvin^{-1}}$ for aluminum this evaluates to a scale of $\qty{4.2}{\milli \kelvin \ \meter^{-1} \ \watt^{-1}}$. For a component of size \qty{1}{\meter} or less where the heat flux is dominated by solar irradiation, the temperature gradients are expected to be no greater than roughly \qty{6}{\kelvin} across the component. 
Our spectral model considers a bimodal temperature distribution. For an RSO with significant thermal gradients, a superior spectral model may be composed of the integral of greybody spectra with temperatures between some $T_C$ and $T_H$. While such a model would have an infinite number of component greybodies, having only the three parameters of the lower and upper temperature bounds $T_C$ and $T_H$ and some apparent intensity of the spectra $\alpha$, the model could still be inferred from measurement in a few number of filters.
\\
\\
Deviations in constant emissivity as a function of wavelength are observed in common spacecraft materials such as aluminum \citep{sainz-menchon_experimental_2022}, Kapton \citep{palm_wavelength_2012}, and silicon solar cells \citep{riverola_mid-infrared_2018} which find variations in the emissivity on the order of \num{0.1} across the thermal IR.
The wavelength and material variations of $f_C$ across the observed IR bands are a second major reason the IR RSO spectra could deviate significantly from a mixture of greybodies. 
While wavelength dependence on scales smaller than IR bandpass widths would largely be integrated out, large scale deviations across the observed IR range on the order of measurement SNR would have a significant effect on the obtained fit. Given this, future work should take effort to model and correct for the wavelength dependence on emissivity; due the greater spectral information, we expect capability to correct for wavelength dependent effects will favour measurement across a greater number of IR bands.
Regardless, correction for this effect is likely less important for observing changes in component temperatures than it is for accurately estimating the absolute temperature of the component.
\\
\\
A caveat of this work is that we consider the scenario where Poissonian noise-sources (such the thermal foreground, shot noise, and sky background) form the dominant source of measurement uncertainty. While this is a major contributing factor for instruments such as the WISE space telescope, particularly for the longer wavelength filters W3 and W4 which are often sky background limited, there are other systematic uncertainties that must be considered working with real observations. For WISE observations, some of these systematic uncertainties include source confusion and contamination from background stars in satellite streaks, background estimation errors, errors in point spread function fitting for elongated satellite streaks, and saturation of sources in the field. Another significant source of uncertainty for the WISE space telescope are calibration errors, such as the uncertainty in the spectral response function of the filters \citep{wright_wide-field_2010}. 
While an error propagation that takes into account these systematic uncertainties is out of the scope of this work, it will be discussed in detail in a future work where we apply these methods to real WISE data.

\subsection{Signal-to-noise ratios and achievable capabilities}
\begin{figure}
    \centering
    \includegraphics[scale=1.0]{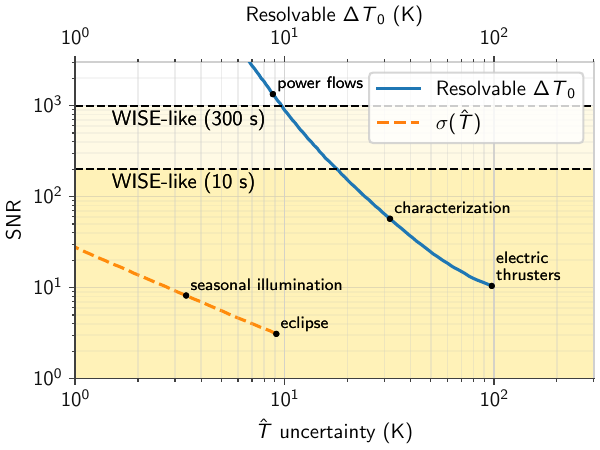}
    \caption{
      RSO observation capabilities which become available at a given SNR assuming Poisson noise dominated observations. The orange series, $\sigma(\hat{T})$, gives the uncertainty in the measurements of the temperature of a single greybody $T_0 = \qty{300}{\kelvin}$ (a single greybody is a more meaningful fit at lower SNR).
      The blue series, resolvable ${\Delta T}_0$, gives the minimum temperature difference between the hot and cold components of the two-component model (eqn. \ref{eqn:2bb}), ${\Delta T}_0$, at which a two-component greybody spectral model can be distinguished from a one-component model at parameters of ${f_C}_0 = 0.7$ and ${T_{\text{eff}}}_0 = \qty{300}{\kelvin}$. To facilitate model preference comparison, both series are based on observations in eight bands logarithmically spaced between the central wavelengths of WISE W1 and W4. The indicative SNR and the available capabilities of a WISE-like space telescope at a given exposure time are provided. Text labels indicate interesting RSO observations capabilities which become available at a given temperature resolution for each series. As the text labels which indicate RSO observation capabilities depend on many external factors, in this plot they serve only as an rough illustration of the capabilities which become available at increasing SNR.
    }
    \label{fig:fontier}
\end{figure}

In this work, our results and discussion have been focused around Poisson noise dominated, WISE-like observations, with total integration times of 10-300s. However, our methodological framework for analysis is relatively flexible to observations using IR telescopes. In \cref{fig:fontier} we conceptually illustrate the RSO observation monitoring capabilities which become available at increasing SNR for a space-based IR SDA telescope. A WISE-like IR space telescope with eight filters and \qty{10}{\second} exposures (not considering any systematic uncertainties in IR photometry) is able to resolve the effective temperature of an RSO well within \qty{1}{\kelvin}. More importantly, resolving an RSO with two greybody components down to a separation of ${\Delta T}_0 = \qty{20}{\kelvin}$ may also become possible. As a result, characterization of RSO based on their component temperatures may become feasible. Future work should explore if characterization of RSO based on their component temperatures permits distinction between active satellites and space debris or rocket bodies unfeasible with a single greybody model.
For a space-based IR telescope approaching Poisson noise dominated observations of an SNR of \num{E3}, the potential of monitoring power flows between components of an RSO also begins to become viable. 
These capabilities are contrasted with the limitations of ground-based telescopes. At these wavelengths atmospheric transmission is low in the thermal IR due to strong absorption by the atmosphere and stringent stray light control and thermal stability of the instrument is required. Therefore, a high sensitivity space telescope unlocks new SDA capabilities from thermal IR spectra unavailable from ground based systems.

\subsection{Future work}
Given the novelty of our two-component spectral model, detailed investigation surrounding the potential capabilities of a space-based IR telescope have yet to be conducted and a full exploration of the information about an RSO discernible from its spectra may still reveal further exciting capabilities. Future works should investigate such capabilities.

To accomplish this, a detailed RSO thermal model (or better, family of models) with order \num{10} or greater components as needed could be built. A model of this scale could describe the temperatures of individual spacecraft systems such as radiators, solar panels, antenna, thrusters, and specific instrumentation and be used to generate a detailed thermal spectra. If this thermal model were coupled with an operational model to describe spacecraft's attitude and the power usage and generation by each component under various operating regimes (station keeping, data processing, battery charging, etc.), the combined model could be used to study methods to deduce operating regimes from the RSO's unresolved spectrum.

As demonstrated, the addition of prior information on ${f_C}_0$ demonstrated a large decrease in the uncertainty of other parameters. Extending this idea, a detailed RSO model would provide the effective areas of each component from every perspective, future study should investigate how, for observations of an RSO where a detailed model is available, this information can best be used to infer the temperatures of components of interest. This study may also explore the power of repeat observations of the same RSO or simultaneous observations from multiple space telescopes, for thermal and operational inference.

In this work the RSO's spectra was assumed to be given by our two-component model (eqn. \ref{eqn:2bb}). While this enabled an analysis of the parameter uncertainty obtained at a given telescope SNR and basic model comparison, a detailed RSO model or serendipitous RSO observations from an existing IR space telescope such as WISE, could be used to further validate the utility of our two-component model and to quantify the magnitude of its limitations such as the wavelength dependence of $f_C$ or component thermal gradients. Our modeling was based on observations at the central filter wavelengths and with simplified observation SNR assumptions. \ref{sec:apdx:broadbands} validates the use of central wavelength sampling for the WISE bands in this paper and estimated that the presented uncertainty in the inferred parameters would be at most about \qty{50}{\percent} larger if the WISE broadband spectral responses were used - acceptable for the conceptual nature of this study and conclusions drawn on the trends and dependencies of this parameter inference.
This being said any space telescope designed specifically for SDA purposes would not be restricted to the four WISE bandpasses an thus need to consider the optimal number of bandpasses, their width, and central wavelengths to minimize this parameter uncertainty. 
It is strongly expected that parameter uncertainty will decrease with a greater number of measurement bandpasses. Yet, as each additional bandpass adds complexity and cost in a telescope design, in a way which is likely to scale superlinearly, an engineering tradeoff between cost, complexity and parameter uncertainty is critically needed to drive future space telescope design studies. 
The precise modeling of the expected SNR for a given bandpass and observed RSO is a course for future work in order to minimize parameter uncertainty. Wider bandpasses collect more flux and hence they will tend to have higher a SNR for the same exposure time. The SNR of the target will also depend strongly on the astrophysical background in the infrared, primarily made up of the Zodiacal light. At the same time, wider bands introduce uncertainty related to the curvature of the spectra within the bandpass and hence creating a tradeoff that needs to be carefully balanced for optimal performance.

Finally, to apply this model to derive intelligence of RSOs using real infrared space telescopes systems, a system-specific error analysis including propagation of systematic uncertainties introduced by the imaging system, data reduction and photometry must be considered in a future work. 

\section{Conclusion}\label{sec:conclusion}

We studied the capability to infer RSO temperatures from unresolved IR spaced-based photometry. 
Following review of the known challenges a single greybody model has for characterizing RSOs and providing insight into their operation, we proposed a two-greybody thermal and spectral model for IR SDA.
Our two-component model represents a conceptual leap over existing models in that it considers multiple characteristic temperatures within a single RSO, yet it is also physically motivated with thermal modeling revealing that due to geometry a bimodal temperature distribution of separations up to \qty{80}{\kelvin} is expected of \textqt{box-wing} type spacecraft designs with compact spacecraft bodies and planar solar panels.
Thermal modeling also revealed this temperature difference to be a strong perspective-invariant indicator of power flows between spacecraft components.

We evaluated the utility of our model for IR SDA by investigating its parameter estimation from photometric observations at wavelengths similar to the WISE IR space telescope bands (\qtylist{3.4; 4.6; 12; 22}{\micro \meter}) and with signal to noise ratios of \num{1000} typical achieved with exposure times ranging \qtyrange{10}{300}{\second} at WISE-like system performance.  

Firstly, we determined the sampling distributions of the model parameters via Monte Carlo simulation. The results of this confirm that all parameters of our spectral model, including the effective RSO temperature, the component temperature difference, and the effective area fraction of the cold greybody are indeed able to be inferred from observations by a thermal IR space telescope to an uncertainty of \qty{0.55}{\kelvin}, \qty{3.5}{\kelvin}, and \num{0.13} respectively for an exemplar RSO in GEO. In general, the effective temperature of the RSO can be precisely recovered while the parameters which resolve the greybodies (the component temperature difference and cold greybody effective area fraction) are recovered with a greater relative uncertainty that increases as the temperature difference decreases.
Our results demonstrate the potential to resolve intra-RSO characteristic temperatures from unresolved thermal IR photometric imaging. 
External constraints such as multiple simultaneous observations, imaging in more IR photometric bands, and particularly prior knowledge on the geometry of the RSO were all found to benefit reducing the uncertainty of model parameter inference.

Secondly, we conducted a model preference study to determine the conditions for which resolving the photometric observations into our two-component model is valid. This study found that for an observation prescription similar to that used for the sampling distribution work (yet with eight logarithmically spaced filters) the two-component model was preferred when the temperature difference was greater than about \qty{15}{\kelvin} and the cold-greybody effective area fraction was not close to $0$ or $1$. Therefore, the studied spatially-unresolved infrared observation of an RSO can robustly resolve the presence of a second greybody component. This provides a novel metric which may aid in characterization of an RSO between active spacecraft, rocket bodies, space debris, and other classes of RSO.

Finally, we outlined the future directions and frontier capabilities of IR SDA. Specifically, study of observation strategies, use of prior information known about an RSO, and optimization of the observation bands would lead to minimizing the parameter uncertainties or SNR requirements. Additionally, the development of a detailed RSO thermal and operational model could be used to study individual component temperatures and capabilities to infer the RSO's onboard operations.

\section*{Acknowledgments}

This research is supported by the Australian Office of National
Intelligence through the National Intelligence and Security Discovery Research Grant award NI230100162. 
This research is supported by an Australian Government Research Training Program (RTP) Scholarship.
This research is partially based on information from ESA DISCOS (Database and Information System Characterising Objects in Space), a single-source reference for launch information, object registration details, launch vehicle descriptions, as well as spacecraft information for all trackable, unclassified objects. We acknowledge ESA's efforts to maintain and operate this database with its APIs.

\bibliographystyle{jasr-model5-names}
\biboptions{authoryear}
\bibliography{
                    refs,
                    zot-sca
                    }

@misc{commonwealth_of_australia_2020_2020,
	title = {2020 {Force} {Structure} {Plan}},
	publisher = {Commonwealth Government of Australia Department of Defence},
	author = {{Australian Department of Defence}},
	year = {2020},
	note = {ISBN: 978-0-9941680-6-1},
}

@ARTICLE{stealth_sat_2022,
       author = {{Sun}, Hanqing and {Qin}, Yuantian},
        title = "{Stealthy Configuration Optimization Design and RCS Characteristics Study of Microsatellite}",
      journal = {Aerospace},
         year = 2022,
        month = dec,
       volume = {9},
       number = {12},
          eid = {815},
        pages = {815},
          doi = {10.3390/aerospace9120815},
       adsurl = {https://ui.adsabs.harvard.edu/abs/2022Aeros...9..815S}
}

@ARTICLE{brammer2008,
       author = {{Brammer}, Gabriel B. and {van Dokkum}, Pieter G. and {Coppi}, Paolo},
        title = "{EAZY: A Fast, Public Photometric Redshift Code}",
      journal = {Astrophysical Journal},
         year = 2008,
        month = oct,
       volume = {686},
       number = {2},
        pages = {1503-1513},
          doi = {10.1086/591786},
archivePrefix = {arXiv},
       eprint = {0807.1533},
 primaryClass = {astro-ph},
       adsurl = {https://ui.adsabs.harvard.edu/abs/2008ApJ...686.1503B}
}

@ARTICLE{HR_langer_2024,
       author = {{Langer}, N. and {Kudritzki}, R.~P.},
        title = "{The spectroscopic Hertzsprung-Russell diagram}",
      journal = {Astronomy and Astrophysics},
         year = 2014,
        month = apr,
       volume = {564},
          eid = {A52},
        pages = {A52},
          doi = {10.1051/0004-6361/201423374},
archivePrefix = {arXiv},
       eprint = {1403.2212},
 primaryClass = {astro-ph.SR},
       adsurl = {https://ui.adsabs.harvard.edu/abs/2014A&A...564A..52L}
}

@ARTICLE{Panter2007,
       author = {{Panter}, Benjamin and {Jimenez}, Raul and {Heavens}, Alan F. and {Charlot}, Stephane},
        title = "{The star formation histories of galaxies in the Sloan Digital Sky Survey}",
      journal = {Monthly Notices of the RAS},
         year = 2007,
        month = jul,
       volume = {378},
       number = {4},
        pages = {1550-1564},
          doi = {10.1111/j.1365-2966.2007.11909.x},
archivePrefix = {arXiv},
       eprint = {astro-ph/0608531},
 primaryClass = {astro-ph},
       adsurl = {https://ui.adsabs.harvard.edu/abs/2007MNRAS.378.1550P}
}

@ARTICLE{Johnson2021,
       author = {{Johnson}, Benjamin D. and {Leja}, Joel and {Conroy}, Charlie and {Speagle}, Joshua S.},
        title = "{Stellar Population Inference with Prospector}",
      journal = {Astrophysics Journal, Supplement},
         year = 2021,
        month = jun,
       volume = {254},
       number = {2},
          eid = {22},
        pages = {22},
          doi = {10.3847/1538-4365/abef67},
archivePrefix = {arXiv},
       eprint = {2012.01426},
 primaryClass = {astro-ph.GA},
       adsurl = {https://ui.adsabs.harvard.edu/abs/2021ApJS..254...22J}
}

@ARTICLE{StellarEv_1967ARAA,
       author = {{Iben}, Jr., Icko},
        title = "{Stellar Evolution Within and off the Main Sequence}",
      journal = {Annual Review of Astronomy and Astrophysics},
         year = 1967,
        month = jan,
       volume = {5},
        pages = {571},
          doi = {10.1146/annurev.aa.05.090167.003035},
       adsurl = {https://ui.adsabs.harvard.edu/abs/1967ARA&A...5..571I}
}

@book{united_nations_office_for_outer_space_affairs_guidelines_2021,
	title = {Guidelines for the {Long}-term {Sustainability} of {Outer} {Space} {Activities} of the {Committee} on the {Peaceful} {Uses} of {Outer} {Space}},
	isbn = {978-92-1-002185-2},
	language = {en},
	publisher = {United Nations},
	author = {{United Nations Office for Outer Space Affairs}},
	month = jan,
        address   = {Vienna, Austria},
	year = {2021},
	doi = {10.18356/9789210021852},
}

@inproceedings{bloom_space_2022,
	address = {Maui, Hawaii},
	title = {Space and {Ground}-{Based} {SDA} {Sensor} {Performance} {Comparisons}},
	language = {en},
	booktitle = {Advanced {Maui} {Optical} and {Space} {Surveillance} {Technologies} {Conference} {Proceedings}},
	author = {Bloom, Amelia and Wysack, Joshua and Griesbach, Jacob D and Lawitzke, Anna},
	year = {2022},
}

@article{johnson_us_1993,
	title = {U.{S}. space surveillance},
	volume = {13},
	issn = {0273-1177},
	doi = {10.1016/0273-1177(93)90563-Q},
	number = {8},
	urldate = {2024-10-27},
	journal = {Advances in Space Research},
	author = {Johnson, Nicholas L.},
	month = aug,
	year = {1993},
	pages = {5--20},
}

@techreport{chatters_space_2009,
	title = {Space {Surveillance} {Network}},
	url = {https://www.jstor.org/stable/resrep13939.26},
	urldate = {2024-10-27},
	institution = {Air University Press},
	author = {Chatters, Edward P. and Crothers, Brian J. and College, Air Command {and} Staff and Seminars, Space Research Electives},
	year = {2009},
	pages = {249--258},
}

@inproceedings{weeden_global_2001,
	title = {{GLOBAL} {SPACE} {SITUATIONAL} {AWARENESS} {SENSORS}},
	booktitle = {Proceedings of the {Advanced} {Maui} {Optical} and {Space} {Surveillance} {Technologies} {Conference}},
	author = {Weeden, Brian and Cefola, Paul and Sankaran, Jaganath},
	year = {2001},
}

@inproceedings{shaddix_daytime_2021,
	title = {Daytime {Optical} {Contributions} {Toward} {Timely} {Space} {Domain} {Awareness} in {Low} {Earth} {Orbit}},
	language = {en},
	author = {Shaddix, Jeﬀ and Key, Cameron and Ferris, Alex and Herring, James and Singh, Navraj and Brost, Todd and Aristoﬀ, Jeﬀ},
	year = {2021},
        address = {Maui, Hawaii},
	booktitle = {Advanced {Maui} {Optical} and {Space} {Surveillance} {Technologies} {Conference} {Proceedings}},
}

@inproceedings{caddy_surprising_2024,
	title = {A {Surprising} {Boost} in {Starlink} {Satellite} {Brightness} at {Optical} {Wavelengths} {During} the {Day}},
	booktitle = {Advanced {Maui} {Optical} and {Space} {Surveillance} {Technologies} {Conference}},
	author = {Caddy, Sarah E. and Spitler, Lee R.},
	month = oct,
	year = {2024},
}

@article{caddy_optical_2024,
	title = {An {Optical} {Daytime} {Astronomy} {Pathfinder} for the {Huntsman} {Telescope}},
	issn = {1323-3580, 1448-6083},
	doi = {10.1017/pasa.2024.43},
	language = {en},
	urldate = {2024-05-20},
	journal = {Publications of the Astronomical Society of Australia},
	author = {Caddy, Sarah E. and Spitler, Lee R. and Ellis, Simon C.},
	month = may,
	year = {2024},
	pages = {1--19},

}

@inproceedings{block_cislunar_2022,
	title = {Cislunar {SDA} with {Low}-{Fidelity} {Sensors} and {Observer} {Uncertaint}},
	booktitle = {Advanced {Maui} {Optical} and {Space} {Surveillance} {Technologies} {Conference}},
	author = {Block, Joshua M and Curtis, H, David and Bettinger, A, Robert and Wilmer, P, Adam},
	year = {2022},
}

@inproceedings{ackermann_systematic_2015,
	title = {A {SYSTEMATIC} {EXAMINATION} {OF} {GROUND}-{BASED} {AND} {SPACE}-{BASED} {APPROACHES} {TO} {OPTICAL} {DETECTION} {AND} {TRACKING} {OF} {SATELLITES}},
	language = {en},
	booktitle = {31st {Space} {Symposium}},
	author = {Ackermann, Mark R and Kiziah, Colonel Rex R and Zimmer, Peter C and McGraw, J T and McGraw, John T and McGraw, J T and Cox, David D},
	month = apr,
	year = {2015},
}

@inproceedings{leitch_sapphire_2010,
	title = {Sapphire: {A} {Small} {Satellite} {System} for the {Surveillance} of {Space}},
	language = {en},
	booktitle = {24th {Annual} {AIAA}/{USU} {Conference} on {Small} {Satellites}},
	author = {Leitch, Robert and Hemphill, Ian},
	year = {2010},
}

@article{du_tentative_2019,
	title = {Tentative design of {SBSS} constellations for {LEO} debris catalog maintenance},
	volume = {155},
	issn = {0094-5765},
	doi = {10.1016/j.actaastro.2018.06.054},
	urldate = {2024-10-28},
	journal = {Acta Astronautica},
	author = {Du, Jianli and Chen, Junyu and Li, Bin and Sang, Jizhang},
	month = feb,
	year = {2019},
	pages = {379--388},
}

@article{gaposchkin_space-based_2000,
	title = {Space-based space surveillance with the {Space}-{Based} {Visible}},
	volume = {23},
	doi = {10.2514/2.4502},
	number = {1},
	journal = {Journal of Guidance, Control, and Dynamics},
	author = {Gaposchkin, E.M. and Von Braun, C. and Sharma, J.},
	year = {2000},
	pages = {148--152},
}

@inproceedings{utzmann_space-based_2014,
	address = {Toronto, Canada},
	title = {Space-based space surveillance and tracking demonstrator: mission and system design},
	copyright = {info:eu-repo/semantics/restrictedAccess},
	isbn = {978-1-63439-986-9},
	shorttitle = {Space-based space surveillance and tracking demonstrator},
	url = {http://www.proceedings.com/25520.html},
	language = {eng},
	urldate = {2024-10-28},
	booktitle = {Utzmann, {J}.; {Wagner}, {A}.; {Silha}, {Jiri}; {Schildknecht}, {Thomas}; {Willemsen}, {P}.; {Teston}, {F}.; {Flohrer}, {T}. (2014). {Space}-based space surveillance and tracking demonstrator: mission and system design. {In}: {Proceedings} of 65th {International} {Astronautical} {Congress} (pp. 1648-1655). {International} {Astronautical} {Federation} ( {IAF} )},
	publisher = {International Astronautical Federation ( IAF )},
	author = {Utzmann, J. and Wagner, A. and Silha, Jiri and Schildknecht, Thomas and Willemsen, P. and Teston, F. and Flohrer, T.},
	collaborator = {Utzmann, J. and Wagner, A. and Silha, Jiri and Schildknecht, Thomas and Willemsen, P. and Teston, F. and Flohrer, T.},
	year = {2014},
	note = {Num Pages: 7},
	pages = {1648--1655},
}

@inproceedings{lee_distinguishing_2017,
	title = {Distinguishing {Active} {Box}-{Wing} and {Cylindrical} {Geostationary} {Satellites} using {IR} {Photometry} with {NASA}'s {WISE} {Spacecraft}},
	language = {en},
	booktitle = {Advanced {Maui} {Optical} and {Space} {Surveillance} {Technologies} {Conference}},
	author = {Lee, Chris H and Seitzer, Patrick and Cutri, Roc M and Grillmair, Carl J and Schildknecht, Thomas and Murray-Krezan, Jeremy J and Bédard, Donald},
	year = {2017},
}

@inproceedings{lee_infrared_2016,
	title = {Infrared {Photometry} of {GEO} {Spacecraft} with {WISE}},
	language = {en},
	booktitle = {Advanced {Maui} {Optical} and {Space} {Surveillance} {Technologies} {Conference}},
	author = {Lee, Chris H and Seitzer, Patrick and Cutri, R M and Grillmair, C J and Schildknecht, T},
	year = {2016},
}

@article{dow_detection_1990,
	title = {The detection of earth orbiting objects by {IRAS}},
	volume = {10},
	issn = {0273-1177},
	url = {https://www.sciencedirect.com/science/article/pii/0273117790903738},
	doi = {10.1016/0273-1177(90)90373-8},
	number = {3},
	urldate = {2024-10-28},
	journal = {Advances in Space Research},
	author = {Dow, Kimberly L. and Sykes, Mark V. and Low, Frank J. and Vilas, Faith},
	month = jan,
	year = {1990},
	pages = {381--384},
}

@inproceedings{hart_quantitative_2014,
	title = {Quantitative {Measurements} of the {Daytime} {Near} {Infrared} {Sky} {Brightness} at the {AEOS} 3.6 m {Telescope}},
	urldate = {2024-10-28},
	booktitle = {Advanced {Maui} {Optical} and {Space} {Surveillance} {Technologies} {Conference}},
	author = {Hart, M. and Jefferies, S. and Hope, D. and Nagy, J. and Williams, S.},
	month = sep,
	year = {2014},
	note = {Conference Name: Advanced Maui Optical and Space Surveillance Technologies Conference
Pages: E73
ADS Bibcode: 2014amos.confE..73H},
}

@inproceedings{hart_resolved_2015,
	title = {Resolved {Observations} of {Geosynchronous} {Satellites} from the 6.5 m {MMT}},
	urldate = {2024-10-28},
	booktitle = {Advanced {Maui} {Optical} and {Space} {Surveillance} {Technologies} {Conference}},
	author = {Hart, M. and Rast, R. and Jefferies, S.},
	month = jan,
	year = {2015},
}

@inproceedings{thomas_ground-based_2019,
	title = {Ground-based, daytime modeling and observations in {SWIR} for satellite custody},
	language = {en},
	author = {Thomas, Grant M and Cobb, Richard G},
	year = {2019},
        booktitle = {Advanced {Maui} {Optical} and {Space} {Surveillance} {Technologies} {Conference}},
}

@inproceedings{howard_geo_2015,
	title = {{GEO} {Collisional} {Risk} {Assessment} {Based} on {Analysis} of {NASA}-{WISE} {Data} and {Modeling}},
	urldate = {2024-10-29},
	booktitle = {Advanced {Maui} {Optical} and {Space} {Surveillance} {Technologies} {Conference}},
	author = {Howard, S. and Murray-Krezan, J. and Dao, P. and Surka, D.},
	month = jan,
	year = {2015},
}

@inproceedings{murakami_infrared_2008,
	title = {The infrared astronomical satellite {AKARI}: overview, highlights of the mission},
	volume = {7010},
	shorttitle = {The infrared astronomical satellite {AKARI}},
	url = {https://www.spiedigitallibrary.org/conference-proceedings-of-spie/7010/70100A/The-infrared-astronomical-satellite-AKARI--overview-highlights-of-the/10.1117/12.788593.full},
	doi = {10.1117/12.788593},
	urldate = {2024-10-29},
	booktitle = {Space {Telescopes} and {Instrumentation} 2008: {Optical}, {Infrared}, and {Millimeter}},
	publisher = {SPIE},
	author = {Murakami, Hiroshi and Matsuhara, Hideo},
	month = jul,
	year = {2008},
	pages = {86--95},
}

@article{neugebauer_infrared_1984,
	title = {The {Infrared} {Astronomical} {Satellite} ({IRAS}) mission.},
	volume = {278},
	issn = {0004-637X},
	url = {https://ui.adsabs.harvard.edu/abs/1984ApJ...278L...1N},
	doi = {10.1086/184209},
	urldate = {2024-10-29},
	journal = {The Astrophysical Journal},
	author = {Neugebauer, G. and Habing, H. J. and van Duinen, R. and Aumann, H. H. and Baud, B. and Beichman, C. A. and Beintema, D. A. and Boggess, N. and Clegg, P. E. and de Jong, T. and Emerson, J. P. and Gautier, T. N. and Gillett, F. C. and Harris, S. and Hauser, M. G. and Houck, J. R. and Jennings, R. E. and Low, F. J. and Marsden, P. L. and Miley, G. and Olnon, F. M. and Pottasch, S. R. and Raimond, E. and Rowan-Robinson, M. and Soifer, B. T. and Walker, R. G. and Wesselius, P. R. and Young, E.},
	month = mar,
	year = {1984},
	note = {Publisher: IOP
ADS Bibcode: 1984ApJ...278L...1N},
	pages = {L1--L6},
}

@inproceedings{gibson_optical-infrared_2013,
	title = {Optical-{Infrared} {Colors} of {GEO} {Satellites}},
	urldate = {2024-10-30},
	booktitle = {Advanced {Maui} {Optical} and {Space} {Surveillance} {Technologies} {Conference}},
	author = {Gibson, B. and Jim, K. and Cognion, R. and Pier, E. A.},
	month = sep,
	year = {2013},
	note = {Conference Name: Advanced Maui Optical and Space Surveillance Technologies Conference
Pages: E67
ADS Bibcode: 2013amos.confE..67G},
}

@article{ellis_case_2008,
	title = {The case for {OH} suppression at near-infrared wavelengths},
	volume = {386},
	issn = {00358711},
	doi = {10.1111/j.1365-2966.2008.13021.x},
	number = {1},
	urldate = {2020-01-16},
	journal = {Monthly Notices of the Royal Astronomical Society},
	author = {Ellis, S. C. and Bland-Hawthorn, J.},
	month = may,
	year = {2008},
	note = {arXiv: 0801.3870},
	pages = {47--64},
}

@incollection{cox_infrared_2002,
	address = {New York, NY},
	title = {Infrared {Astronomy}},
	copyright = {https://www.springernature.com/gp/researchers/text-and-data-mining},
	isbn = {978-1-4612-7037-9 978-1-4612-1186-0},
	url = {https://link.springer.com/10.1007/978-1-4612-1186-0_7},
	language = {en},
	urldate = {2024-10-31},
	booktitle = {Allen’s {Astrophysical} {Quantities}},
	publisher = {Springer New York},
	author = {Tokunaga, A. T.},
	editor = {Cox, Arthur N.},
	year = {2002},
	doi = {10.1007/978-1-4612-1186-0_7},
	pages = {143--167},
}

@article{gehrz_nasa_2007,
	title = {The {NASA} {Spitzer} {Space} {Telescope}},
	volume = {78},
	issn = {0034-6748},
	doi = {10.1063/1.2431313},
	number = {1},
	urldate = {2024-10-31},
	journal = {Review of Scientific Instruments},
	author = {Gehrz, R. D. and Roellig, T. L. and Werner, M. W. and Fazio, G. G. and Houck, J. R. and Low, F. J. and Rieke, G. H. and Soifer, B. T. and Levine, D. A. and Romana, E. A.},
	month = jan,
	year = {2007},
	pages = {011302},
}

@inproceedings{glasse_infrared_1993,
	title = {Infrared {Measurements} of the {Sky} {Noise} {Power} {Spectrum} at {Mauna} {Kea}},
	volume = {41},
	url = {https://ui.adsabs.harvard.edu/abs/1993ASPC...41..393G},
	urldate = {2024-10-31},
	booktitle = {Astronomical {Infrared} {Spectroscopy}. {Future} {Observational} {Directions}},
	publisher = {Astronomical Society of the Pacific},
	author = {Glasse, A. C. H. and Casali, M. M.},
	month = jan,
	year = {1993},
	note = {Conference Name: Astronomical Infrared Spectroscopy: Future Observational Directions
ADS Bibcode: 1993ASPC...41..393G},
	pages = {393},
}

@misc{sejba_space_2023,
	title = {Space {Domain} {Awareness}: {Doctrine} for {Space} {Forces}},
	publisher = {United States Space Force},
	author = {Sejba, Timothy},
	month = nov,
	year = {2023},
}

@article{baker_comprehensive_2024,
	title = {A comprehensive review on {Cislunar} expansion and space domain awareness},
	volume = {147},
	issn = {0376-0421},
	doi = {10.1016/j.paerosci.2024.101019},
	urldate = {2024-12-02},
	journal = {Progress in Aerospace Sciences},
	author = {Baker-McEvilly, Brian and Bhadauria, Surabhi and Canales, David and Frueh, Carolin},
	month = may,
	year = {2024},
	pages = {101019},
}

@article{bertin_sextractor_1996,
	title = {{SExtractor}: {Software} for source extraction.},
	volume = {117},
	issn = {0365-01380004-6361},
	shorttitle = {{SExtractor}},
	doi = {10.1051/aas:1996164},
	journal = {Astronomy and Astrophysics Supplement Series},
	author = {Bertin, E. and Arnouts, S.},
	month = jun,
	year = {1996},
	note = {ADS Bibcode: 1996A\&AS..117..393B},
	pages = {393--404},
}

@software{larry_bradley_2024_13989456,
  author       = {Larry Bradley and
                  Brigitta Sip{\H o}cz and
                  Thomas Robitaille and
                  Erik Tollerud and
                  Z\`e Vin{\'{\i}}cius and
                  Christoph Deil and
                  Kyle Barbary and
                  Tom J Wilson and
                  Ivo Busko and
                  Axel Donath and
                  Hans Moritz G{\"u}nther and
                  Mihai Cara and
                  P. L. Lim and
                  Sebastian Me{\ss}linger and
                  Simon Conseil and
                  Zach Burnett and
                  Azalee Bostroem and
                  Michael Droettboom and
                  E. M. Bray and
                  Lars Andersen Bratholm and
                  Adam Ginsburg and
                  William Jamieson and
                  Geert Barentsen and
                  Matt Craig and
                  Brett M. Morris and
                  Marshall Perrin and
                  Shivangee Rathi and
                  Sergio Pascual and
                  Iskren Y. Georgiev},
  title        = {astropy/photutils: 2.0.2},
  month        = oct,
  year         = 2024,
  publisher    = {Zenodo},
  version      = {2.0.2},
  doi          = {10.5281/zenodo.13989456},
  url          = {https://doi.org/10.5281/zenodo.13989456},
}

@article{collaboration_astropy_2022,
	title = {The {Astropy} {Project}: {Sustaining} and {Growing} a {Community}-oriented {Open}-source {Project} and the {Latest} {Major} {Release} (v5.0) of the {Core} {Package}},
	volume = {935},
	issn = {0004-637X, 1538-4357},
	shorttitle = {The {Astropy} {Project}},
	url = {http://arxiv.org/abs/2206.14220},
	doi = {10.3847/1538-4357/ac7c74},
	number = {2},
	urldate = {2025-07-21},
	journal = {The Astrophysical Journal},
	author = {Collaboration, The Astropy and Price-Whelan, Adrian M. and Lim, Pey Lian and Earl, Nicholas and Starkman, Nathaniel and Bradley, Larry and Shupe, David L. and Patil, Aarya A. and Corrales, Lia and Brasseur, C. E. and Nöthe, Maximilian and Donath, Axel and Tollerud, Erik and Morris, Brett M. and Ginsburg, Adam and Vaher, Eero and Weaver, Benjamin A. and Tocknell, James and Jamieson, William and Kerkwijk, Marten H. van and Robitaille, Thomas P. and Merry, Bruce and Bachetti, Matteo and Günther, H. Moritz and Aldcroft, Thomas L. and Alvarado-Montes, Jaime A. and Archibald, Anne M. and Bódi, Attila and Bapat, Shreyas and Barentsen, Geert and Bazán, Juanjo and Biswas, Manish and Boquien, Médéric and Burke, D. J. and Cara, Daria and Cara, Mihai and Conroy, Kyle E. and Conseil, Simon and Craig, Matthew W. and Cross, Robert M. and Cruz, Kelle L. and D'Eugenio, Francesco and Dencheva, Nadia and Devillepoix, Hadrien A. R. and Dietrich, Jörg P. and Eigenbrot, Arthur Davis and Erben, Thomas and Ferreira, Leonardo and Foreman-Mackey, Daniel and Fox, Ryan and Freij, Nabil and Garg, Suyog and Geda, Robel and Glattly, Lauren and Gondhalekar, Yash and Gordon, Karl D. and Grant, David and Greenfield, Perry and Groener, Austen M. and Guest, Steve and Gurovich, Sebastian and Handberg, Rasmus and Hart, Akeem and Hatfield-Dodds, Zac and Homeier, Derek and Hosseinzadeh, Griffin and Jenness, Tim and Jones, Craig K. and Joseph, Prajwel and Kalmbach, J. Bryce and Karamehmetoglu, Emir and Kałuszyński, Mikołaj and Kelley, Michael S. P. and Kern, Nicholas and Kerzendorf, Wolfgang E. and Koch, Eric W. and Kulumani, Shankar and Lee, Antony and Ly, Chun and Ma, Zhiyuan and MacBride, Conor and Maljaars, Jakob M. and Muna, Demitri and Murphy, N. A. and Norman, Henrik and O'Steen, Richard and Oman, Kyle A. and Pacifici, Camilla and Pascual, Sergio and Pascual-Granado, J. and Patil, Rohit R. and Perren, Gabriel I. and Pickering, Timothy E. and Rastogi, Tanuj and Roulston, Benjamin R. and Ryan, Daniel F. and Rykoff, Eli S. and Sabater, Jose and Sakurikar, Parikshit and Salgado, Jesús and Sanghi, Aniket and Saunders, Nicholas and Savchenko, Volodymyr and Schwardt, Ludwig and Seifert-Eckert, Michael and Shih, Albert Y. and Jain, Anany Shrey and Shukla, Gyanendra and Sick, Jonathan and Simpson, Chris and Singanamalla, Sudheesh and Singer, Leo P. and Singhal, Jaladh and Sinha, Manodeep and Sipőcz, Brigitta M. and Spitler, Lee R. and Stansby, David and Streicher, Ole and Šumak, Jani and Swinbank, John D. and Taranu, Dan S. and Tewary, Nikita and Tremblay, Grant R. and Val-Borro, Miguel de and Kooten, Samuel J. Van and Vasović, Zlatan and Verma, Shresth and Cardoso, José Vinícius de Miranda and Williams, Peter K. G. and Wilson, Tom J. and Winkel, Benjamin and Wood-Vasey, W. M. and Xue, Rui and Yoachim, Peter and ZHANG, Chen and Zonca, Andrea},
	month = aug,
	year = {2022},
	note = {arXiv:2206.14220 [astro-ph]},
	pages = {167},
}

@article{opencv_library,
    author = {Bradski, G.},
    journal = {Dr. Dobb's Journal of Software Tools},
    title = {{The OpenCV Library}},
    year = {2000}
}

@article{ceplecha_meteor_1998,
  title = {Meteor {{Phenomena}} and {{Bodies}}},
  author = {Ceplecha, Zden{\v e}k and Borovi{\v c}ka, Ji{\v r}{\'I} and Elford, W. Graham and ReVelle, Douglas O. and Hawkes, Robert L. and Porub{\v c}an, Vladim{\'I}r and {\v S}imek, Milo{\v s}},
  year = {1998},
  month = sep,
  journal = {Space Science Reviews},
  volume = {84},
  number = {3},
  pages = {327--471},
  issn = {1572-9672},
  doi = {10.1023/A:1005069928850},
  url = {https://doi.org/10.1023/A:1005069928850},
  urldate = {2025-09-23},
  langid = {english}
}

@book{gilmore_spacecraft_2002,
  title = {Spacecraft Thermal Control Handbook. {{Volume I}}, {{Fundamental}} Technologies},
  author = {Gilmore, David},
  year = {2002},
  month = jan,
  edition = {2nd ed.},
  publisher = {Aerospace Press},
  url = {https://research.ebsco.com/linkprocessor/plink?id=22e2bcb6-4ced-3f0f-ac0b-14f936f0ba5b},
  urldate = {2025-09-24},
  isbn = {978-1-60119-203-5},
  langid = {english}
}

@article{kim_analytical_2010,
  title = {Analytical and Numerical Approaches of a Solar Array Thermal Analysis in a Low-Earth Orbit Satellite},
  author = {Kim, Hui Kyung and Han, Cho Young},
  year = {2010},
  month = dec,
  journal = {Advances in Space Research},
  volume = {46},
  number = {11},
  pages = {1427--1439},
  issn = {0273-1177},
  doi = {10.1016/j.asr.2010.08.023},
  url = {https://www.sciencedirect.com/science/article/pii/S0273117710005818},
  urldate = {2025-09-24}
}

@inproceedings{allworth_use_2024,
  title = {The {{Use}} of {{Flyby Space-to-Space Non-Earth Imagery}} to {{Rapidly Identify}} and {{Characterise Unknown Objects}}},
  booktitle = {Advanced {{Maui Optical}} and {{Space Surveillance Technologies Conference}}},
  author = {Allworth, James and Bartlett, Stuart and Kirkwood, Samantha and Vincent, Karla and Dawe, Hannah and Sines, Jack and Harris, Toby and Jayakody, Hiranya and Crowe, William},
  year = {2024},
  address = {Maui, Hawai'i},
  langid = {english}
}

@article{bruzual_stellar_2003,
  title = {Stellar Population Synthesis at the Resolution of 2003},
  author = {Bruzual, G. and Charlot, S.},
  year = {2003},
  month = oct,
  journal = {Monthly Notices of the Royal Astronomical Society},
  volume = {344},
  pages = {1000--1028},
  publisher = {OUP},
  issn = {0035-8711},
  doi = {10.1046/j.1365-8711.2003.06897.x},
  url = {https://ui.adsabs.harvard.edu/abs/2003MNRAS.344.1000B},
  urldate = {2024-12-10}
}

@article{cook_augmented_2025,
  title = {Augmented {{Illumination}} of {{Resident Space Objects}} in {{Selected Lagrange-Point Orbits}} via {{Space-Based Mirrors Conducting Proximity Operations}}},
  author = {Cook, Alec E. and Bettinger, Robert A. and Dahlke, Jacob A.},
  year = {2025},
  month = apr,
  journal = {Aerotecnica Missili \& Spazio},
  issn = {2524-6968},
  doi = {10.1007/s42496-025-00257-5},
  url = {https://doi.org/10.1007/s42496-025-00257-5},
  urldate = {2025-04-10},
  langid = {english}
}

@article{cooke_classification_2025,
  title = {Classification of {{LEO}} Satellites Using Occultations of Background Stars},
  author = {Cooke, Benjamin F and Pollacco, Don and Blake, James A and Chote, Paul and Eves, Stuart and Feline, Will and Privett, Grant},
  year = {2025},
  month = jan,
  journal = {RAS Techniques and Instruments},
  volume = {4},
  pages = {rzaf016},
  issn = {2752-8200},
  doi = {10.1093/rasti/rzaf016},
  url = {https://doi.org/10.1093/rasti/rzaf016},
  urldate = {2025-05-23}
}

@article{fitzmaurice_detection_2021,
  title = {Detection and Correlation of Geosynchronous Objects in {{NASA}}'s {{Wide-field Infrared Survey Explorer}} Images},
  author = {Fitzmaurice, Joshua and B{\'e}dard, Donald and Lee, Chris H. and Seitzer, Patrick},
  year = {2021},
  month = jun,
  journal = {Acta Astronautica},
  volume = {183},
  pages = {176--198},
  issn = {0094-5765},
  doi = {10.1016/j.actaastro.2021.03.009},
  url = {https://www.sciencedirect.com/science/article/pii/S0094576521001211},
  urldate = {2024-07-02}
}

@phdthesis{fitzmaurice_streak_2018,
  type = {M.{{Sc}}},
  title = {{{STREAK DETECTION AND PHOTOMETRY WITH THE NASA WIDE-FIELD INFRARED SURVEY EXPLORER}}},
  author = {Fitzmaurice, Joshua},
  year = {2018},
  month = may,
  langid = {english},
  school = {Royal Military College of Canada}
}

@techreport{gaposchkin_infrared_1995,
  title = {Infrared {{Detections}} of {{Satellites}} with {{IRAS}}:},
  shorttitle = {Infrared {{Detections}} of {{Satellites}} with {{IRAS}}},
  author = {Gaposchkin, E. M. and Bergemann, R. J.},
  year = {1995},
  month = sep,
  address = {Fort Belvoir, VA},
  institution = {Defense Technical Information Center},
  doi = {10.21236/ADA537531},
  url = {http://www.dtic.mil/docs/citations/ADA537531},
  urldate = {2024-10-30},
  langid = {english}
}

@book{huterer_course_2023,
  title = {A Course in Cosmology: From Theory to Practice},
  shorttitle = {A Course in Cosmology},
  author = {Huterer, Dragan},
  year = {2023},
  publisher = {Cambridge University Press},
  address = {Cambridge, United Kingdom ; New York, NY, USA},
  isbn = {978-1-316-51359-0},
  lccn = {QB981 .H875 2023}
}

@inproceedings{kielkopf_remote_2022,
  title = {Remote {{Sensing}} of {{Satellite Activity}} through {{Optical}} and {{Infrared Temporal Differential Spectrophotometry Informed}} by {{Analysis}} of {{Noise}}},
  booktitle = {Advanced {{Maui Optical}} and {{Space Surveillance Technologies}} Conference},
  author = {Kielkopf, John and Clark, Frank O},
  year = {2022},
  address = {Maui, Hawai'i},
  langid = {english}
}

@inproceedings{mccall_thermal_2013,
  title = {Thermal Modeling of Space Debris via {{Finite Element Analysis}}},
  booktitle = {Thermal Modeling of Space Debris via {{Finite Element Analysis}}},
  author = {McCall, Paul and Sharples, Rachel},
  year = {2013},
  address = {Maui, Hawai'i},
  langid = {english}
}

@article{miller_new_2007,
  title = {A {{New Sensor Allocation Algorithm}} for the {{Space Surveillance Network}}},
  author = {Miller, James G. (Gil)},
  year = {2007},
  journal = {Military Operations Research},
  volume = {12},
  number = {1},
  eprint = {43941064},
  eprinttype = {jstor},
  pages = {57--70},
  publisher = {Military Operations Research Society},
  issn = {1082-5983},
  url = {https://www.jstor.org/stable/43941064},
  urldate = {2024-09-18}
}

@inproceedings{myers_hall_2016,
  title = {Hall {{Thruster Thermal Modeling}} and {{Test Data Correlation}}},
  booktitle = {52nd {{AIAA}}/{{SAE}}/{{ASEE Joint Propulsion Conference}}},
  author = {Myers, James L. and Kamhawi, Hani and Yim, John and Clayman, Layren},
  year = {2016},
  month = jul,
  publisher = {{American Institute of Aeronautics and Astronautics}},
  address = {Salt Lake City, UT},
  doi = {10.2514/6.2016-4535},
  url = {http://arc.aiaa.org/doi/10.2514/6.2016-4535},
  urldate = {2024-09-25},
  isbn = {978-1-62410-406-0},
  langid = {english}
}

@inproceedings{oconnell_concept_2024,
  title = {Concept of {{Operation}} and {{Initial Performance Summary}} of the {{NorthStar Space-Based Optical SSA System}}},
  booktitle = {Advanced {{Maui Optical}} and {{Space Surveillance Technologies Conference}}},
  author = {O'Connell, Daniel and Leclerc, Jean-Claude and Giammichele, Noemi and Comeau, Ryan and Gollu, Narendra and Phou, Naron and Pelletier, Frederic},
  year = {2024},
  address = {Maui, Hawai'i},
  langid = {english}
}

@article{palm_wavelength_2012,
  title = {Wavelength and Temperature Dependence of Continuous-Wave Laser Absorptance in {{Kapton}}{\textsuperscript{{\textregistered}}} Thin Films},
  author = {Palm, William J.},
  year = {2012},
  month = jul,
  journal = {Optical Engineering},
  volume = {51},
  number = {12},
  pages = {121802},
  issn = {0091-3286},
  doi = {10.1117/1.OE.51.12.121802},
  url = {http://opticalengineering.spiedigitallibrary.org/article.aspx?doi=10.1117/1.OE.51.12.121802},
  urldate = {2025-01-24},
  langid = {english}
}

@techreport{paxson_space_2008,
  title = {Space {{Object Temperature Determination}} from {{Multi-Band Infrared Measurements}}:},
  shorttitle = {Space {{Object Temperature Determination}} from {{Multi-Band Infrared Measurements}}},
  author = {Paxson, Charles and Snell, Hilary E. and Griffin, James M. and Kraemer, Kathleen and Price, Steve and Kendra, Mike and Mizuno, Don},
  year = {2008},
  month = sep,
  address = {Fort Belvoir, VA},
  institution = {Defense Technical Information Center},
  doi = {10.21236/ADA501247},
  url = {http://www.dtic.mil/docs/citations/ADA501247},
  urldate = {2024-10-10},
  langid = {english}
}

@inproceedings{pearce_rapid_2021,
  title = {Rapid {{Discrimination}} of {{Resident Space Objects Using Near-Infrared Photometry}}},
  booktitle = {Advanced {{Maui Optical}} and {{Space Surveillance Technologies}} Conference},
  author = {Pearce, Eric C and Krantz, Harrison and Block, Adam and Sease, Brad and Kirshner, Mitchell},
  year = {2021},
  address = {Maui, Hawai'i},
  langid = {english}
}

@article{reyes_analysis_2023,
  title = {Analysis of {{Spacecraft Materials Discrimination Using Color Indices}} for {{Remote Sensing}} for {{Space Situational Awareness}}},
  author = {Reyes, Jacqueline A. and Cowardin, Heather M. and {Velez-Reyes}, Miguel},
  year = {2023},
  month = sep,
  journal = {The Journal of the Astronautical Sciences},
  volume = {70},
  number = {5},
  pages = {33},
  issn = {2195-0571},
  doi = {10.1007/s40295-023-00400-z},
  url = {https://doi.org/10.1007/s40295-023-00400-z},
  urldate = {2024-10-28},
  langid = {english}
}

@article{riverola_mid-infrared_2018,
  title = {Mid-Infrared Emissivity of Crystalline Silicon Solar Cells},
  author = {Riverola, A. and Mellor, A. and Alonso Alvarez, D. and Ferre Llin, L. and Guarracino, I. and Markides, C. N. and Paul, D. J. and Chemisana, D. and {Ekins-Daukes}, N.},
  year = {2018},
  month = jan,
  journal = {Solar Energy Materials and Solar Cells},
  volume = {174},
  pages = {607--615},
  issn = {0927-0248},
  doi = {10.1016/j.solmat.2017.10.002},
  url = {https://www.sciencedirect.com/science/article/pii/S0927024817305500},
  urldate = {2024-10-30}
}

@article{sainz-menchon_experimental_2022,
  title = {Experimental and Numerical Study of the Emissivity of Rolled Aluminum},
  author = {{Sainz-Mench{\'o}n}, M. and {Gabirondo-L{\'o}pez}, J. and {Gonz{\'a}lez de Arrieta}, I. and Ech{\'a}niz, T. and L{\'o}pez, G. A.},
  year = {2022},
  month = dec,
  journal = {Infrared Physics \& Technology},
  volume = {127},
  pages = {104380},
  issn = {1350-4495},
  doi = {10.1016/j.infrared.2022.104380},
  url = {https://www.sciencedirect.com/science/article/pii/S1350449522003619},
  urldate = {2024-10-30}
}

@inproceedings{skinner_ir_2007,
  title = {{{IR Spectrophotometric Observations}} of {{Geosynchronous Satellites}}},
  booktitle = {Advanced {{Maui Optical}} and {{Space Surveillance Technologies}} Conference},
  author = {Skinner, M. A. and Payne, T. and Russell, R. and Gutierrez, D. and Crawford, K. and Harrington, D. and Kim, David S. and Lynch, D. and Rudy, R.},
  year = {2007},
  address = {Maui, Hawai'i},
  url = {https://www.semanticscholar.org/paper/IR-Spectrophotometric-Observations-of-Satellites-Skinner-Payne/4fef1f27515c76476e6eacbbb0ba821d73792b1d},
  urldate = {2024-11-08}
}

@article{skinner_observations_2014,
  title = {Observations in the Thermal {{IR}} and Visible of a Retired Satellite in the Graveyard Orbit, and Comparisons to Active Satellites in {{GEO}}},
  author = {Skinner, Mark A. and Russell, Ray W. and Kelecy, Tom and Gregory, Steve and Rudy, Richard J. and Kim, Daryl L. and Crawford, Kirk},
  year = {2014},
  month = dec,
  journal = {Acta Astronautica},
  volume = {105},
  number = {1},
  pages = {1--10},
  issn = {0094-5765},
  doi = {10.1016/j.actaastro.2014.08.016},
  url = {https://www.sciencedirect.com/science/article/pii/S0094576514003233},
  urldate = {2025-05-10}
}

@inproceedings{wright_infrared_2023,
  title = {Infrared {{Sensing}} for {{Space-Based Space Domain Awareness}}},
  booktitle = {Advanced {{Maui Optical}} and {{Space Surveillance Technologies Conference}}},
  author = {Wright, Raymond H and Gordon, Michael and Cleve, Jeffery E Van and Schroots, Hans},
  year = {2023},
  address = {Maui, Hawai'i},
  langid = {english}
}

@article{wright_wide-field_2010,
  title = {The {{Wide-field Infrared Survey Explorer}} ({{WISE}}): {{Mission Description}} and {{Initial On-orbit Performance}}},
  shorttitle = {The {{Wide-field Infrared Survey Explorer}} ({{WISE}})},
  author = {Wright, Edward L. and Eisenhardt, Peter R. M. and Mainzer, Amy K. and Ressler, Michael E. and Cutri, Roc M. and Jarrett, Thomas and Kirkpatrick, J. Davy and Padgett, Deborah and McMillan, Robert S. and Skrutskie, Michael and Stanford, S. A. and Cohen, Martin and Walker, Russell G. and Mather, John C. and Leisawitz, David and Gautier, III, Thomas N. and McLean, Ian and Benford, Dominic and Lonsdale, Carol J. and Blain, Andrew and Mendez, Bryan and Irace, William R. and Duval, Valerie and Liu, Fengchuan and Royer, Don and Heinrichsen, Ingolf and Howard, Joan and Shannon, Mark and Kendall, Martha and Walsh, Amy L. and Larsen, Mark and Cardon, Joel G. and Schick, Scott and Schwalm, Mark and Abid, Mohamed and Fabinsky, Beth and Naes, Larry and Tsai, Chao-Wei},
  year = {2010},
  month = dec,
  journal = {The Astronomical Journal},
  volume = {140},
  pages = {1868--1881},
  publisher = {IOP},
  issn = {0004-6256},
  doi = {10.1088/0004-6256/140/6/1868},
  url = {https://ui.adsabs.harvard.edu/abs/2010AJ....140.1868W},
  urldate = {2025-03-11}
}

@article{houck_infrared_2004,
  title = {The {{Infrared Spectrograph}}* ({{IRS}}) on the {{Spitzer Space Telescope}}},
  author = {Houck, J. R. and Roellig, T. L. and {van Cleve}, J. and Forrest, W. J. and Herter, T. and Lawrence, C. R. and Matthews, K. and Reitsema, H. J. and Soifer, B. T. and Watson, D. M. and Weedman, D. and Huisjen, M. and Troeltzsch, J. and Barry, D. J. and {Bernard-Salas}, J. and Blacken, C. E. and Brandl, B. R. and Charmandaris, V. and Devost, D. and Gull, G. E. and Hall, P. and Henderson, C. P. and Higdon, S. J. U. and Pirger, B. E. and Schoenwald, J. and Sloan, G. C. and Uchida, K. I. and Appleton, P. N. and Armus, L. and Burgdorf, M. J. and {Fajardo-Acosta}, S. B. and Grillmair, C. J. and Ingalls, J. G. and Morris, P. W. and Teplitz, H. I.},
  year = {2004},
  month = sep,
  journal = {The Astrophysical Journal Supplement Series},
  volume = {154},
  number = {1},
  pages = {18},
  issn = {0067-0049},
  doi = {10.1086/423134},
  url = {https://dx.doi.org/10.1086/423134},
  urldate = {2025-06-14},
  langid = {english}
}

@article{kendrew_mid-infrared_2015,
  title = {The {{Mid-Infrared Instrument}} for the {{{\emph{James Webb Space Telescope}}}} , {{IV}}: {{The Low-Resolution Spectrometer}}},
  shorttitle = {The {{Mid-Infrared Instrument}} for the {{{\emph{James Webb Space Telescope}}}} , {{IV}}},
  author = {Kendrew, Sarah and Scheithauer, Silvia and Bouchet, Patrice and Amiaux, Jerome and Azzollini, Ruym{\'a}n and Bouwman, Jeroen and Chen, C. H. and Dubreuil, D. and Fischer, Sebastian and Glasse, Alistair and Greene, T. P. and Lagage, P.-O. and Lahuis, Fred and Ronayette, Samuel and Wright, David and Wright, G. S.},
  year = {2015},
  month = jul,
  journal = {Publications of the Astronomical Society of the Pacific},
  volume = {127},
  number = {953},
  pages = {623--632},
  issn = {00046280, 15383873},
  doi = {10.1086/682255},
  url = {http://iopscience.iop.org/article/10.1086/682255},
  urldate = {2025-06-14},
  langid = {english}
}

@article{mason_cubesat_2018,
  title = {{{CubeSat On-Orbit Temperature Comparison}} to {{Thermal-Balance-Tuned-Model Predictions}}},
  author = {Mason, James Paul and Lamprecht, Bret and Woods, Thomas N. and Downs, Chloe},
  year = {2018},
  month = jan,
  journal = {Journal of Thermophysics and Heat Transfer},
  volume = {32},
  number = {1},
  pages = {237--255},
  issn = {0887-8722, 1533-6808},
  doi = {10.2514/1.T5169},
  url = {https://arc.aiaa.org/doi/10.2514/1.T5169},
  urldate = {2024-12-22},
  langid = {english}
}

@article{blanchete_thermal_2024,
  title = {Thermal Design, Analysis and Test: {{Framework}} for {{CubeSat}} in Low {{Earth}} Orbit},
  shorttitle = {Thermal Design, Analysis and Test},
  author = {Blanchete, Narimane and Bah, Abdellah},
  year = {2024},
  month = dec,
  journal = {Results in Engineering},
  volume = {24},
  pages = {103401},
  issn = {2590-1230},
  doi = {10.1016/j.rineng.2024.103401},
  url = {https://www.sciencedirect.com/science/article/pii/S2590123024016530},
  urldate = {2025-06-14}
}

@article{nino_hst_2008,
  title = {Hst Focus Variations with Temperatures},
  author = {Nino, D and Makidon, R and Lallo, M and Sahu, K and Sirianni, M and Casertano, S},
  year = {2008},
  journal = {Instrument Science Report ACS},
  volume = {3}
}

@article{liu_dynamic_2019,
  title = {Dynamic Characteristics of Flexible Spacecraft with Double Solar Panels Subjected to Solar Radiation},
  author = {Liu, Lun and Wang, Xiaodong and Sun, Shupeng and Cao, Dengqing and Liu, Xiyu},
  year = {2019},
  month = feb,
  journal = {International Journal of Mechanical Sciences},
  volume = {151},
  pages = {22--32},
  issn = {0020-7403},
  doi = {10.1016/j.ijmecsci.2018.10.067},
  url = {https://www.sciencedirect.com/science/article/pii/S0020740318315248},
  urldate = {2025-06-25}
}

\appendix
\section{Photometric analysis of WISE data}\label{sec:apdx:snr}
In order to determine a realistic SNR that can be expected of an observation of a geosynchronous satellite by an infrared space telescope, we use images from data which have been retrieved from the NASA Infrared Science Archive\footnote{\url{https://irsa.ipac.caltech.edu/applications/wise/}}. We note that we do not consider any systematic uncertainties introduce in WISE data. Detailed modelling of systematic uncertainties in WISE data are out of the scope of this work, and are being considered in a follow up paper. High level co-added data products have had satellites removed in processing, so we choose to use the WISE Level-1b data products from the single exposure catalogue. These images have had preliminary instrumental, astrometric and photometric calibrations applied by the WISE data pipeline and can be flux calibrated using the {\texttt{MAGZP}} header keyword\footnote{\url{https://wise2.ipac.caltech.edu/docs/release/allsky/expsup/sec4_4h.html}}. 

Following the results of \cite{fitzmaurice_detection_2021, lee_distinguishing_2017, lee_infrared_2016}, satellites are identified in the W3 images (with the highest SNR) by searching for elongated streaks in WISE data characteristic of geosynchronous satellites moving with respect to the background stars during the exposure. Following \cite{fitzmaurice_detection_2021} we use the Hough and Canny edge detection transforms for this purpose from the python implementation of \texttt{OpenCV} \citep{opencv_library} as a robust first pass for satellite source detection. For this analysis we choose the image of Echostar 4 (NORAD catalogue ID: 25331) shown in \cref{fig:wise_images} (image catalogue ID: 01842a060). The image has been matched to Echostar 4 by comparing the {\texttt{astropy}} \citep{collaboration_astropy_2022} computed angular separation of the midpoint of the satellite streak using the WCS (World Coordinate System) information in the image header, to the catalogue position of Geosychronus satellites using historical TLE (Two Line Element) records. These are retrieved from Space Track \footnote{\url{https://www.space-track.org/auth/login}}. The predicted satellite location can be computed using {\texttt{skyfield}} based on the image acquisition time {\texttt{DATE-OBS}}, the geographic longitude {\texttt{GEOLON}}, latitude {\texttt{GEOLAT}} and altitude {\texttt{GEOALT}} of the WISE space telescope at the time of image acquisition found in the image header. 

To determine the SNR of the observation, we use the python implementation of {\texttt{Source Extractor}} \citep{bertin_sextractor_1996} to derive elliptical aperture and sky annulus parameters to measure the total flux of the source and the sky background respectively. This is applied for satellite sources in an image where $1-\frac{b}{a} > 0.8$ where b and a are the semi minor and semi major axis of the elliptical aperture respectively and the detection threshold is set to $1.1 \times$ the global sky background RMS derived from the {\texttt{Source Extractor}} computed background. There are no standard elliptical aperture corrections for WISE data, so these are derived for each filter using the chosen image. {\texttt{astropy photutils}} \citep{larry_bradley_2024_13989456} is used to apply the apertures to the data. The source aperture parameters used are $a = 90$ pixels, and $b = 10$ pixels with a $\theta$ taken from the {\texttt{Source Extractor}} elliptical source finder. For the background we use (a + 10, a + 20) pixels and (b + 10, b + 20) pixels for the inner and outer apertures respectively. We also employ bright star and bad pixel masking to ensure an accurate background estimate. Bright stars for which $SNR > 5$ are detected using {\texttt{Source Extractor}} and masked by a simple 2D Gaussian point spread function mask scaled by the source flux. Pixels in the affected region are replaced using the local background median. Bad pixel masks are also implemented for nan values which are introduced by the WISE pipeline. Following \cite{fitzmaurice_detection_2021}, bad pixels are in-painted. In this work we use {\texttt{astropy}}'s {\texttt{interpolate\_replace\_nans}. 

The SNR measured for filters W1 - W4 are found to be 181, 430, 1203, 206 respectively. The SNR is generally highest in W3 because the central wavelength of the filter is located on the peak of the SED for an object at \qty{300}{\kelvin}, a typical temperature for a geosynchronous satellite. WISE survey images are all limited in exposure time to \qty{7.7}{\second} for W1 and W2, and \qty{8.8}{\second} for W3 and W4, however a dedicated SDA facility would not be constrained by the limitation of an all sky survey, and so may spend more time per target. To estimate what SNR could be achieved in a reasonable exposure time, $t_{\text{exp}}$, (or total time per target for stacked exposures) we scale the SNR's found for WISE images by $\sqrt{t_{\text{exp}}}$. This scaling is valid in the sky background limited regime and assuming both source and sky signal are not varying and accumulate linearly with exposure time. The sky background limited regime is appropriate for WISE observations due to the intensity of the Zodiacal Light sky background in WISE bandpasses. Using this method we find that an $\text{SNR} >\num{1000}$ can be achieved in all filters with an exposure time of at least \qty{300}{\second}. A caveat of this is that higher SNR observations produced with longer exposure times may still carry systematic uncertainties. A full analysis of the systematic uncertainties introduced in WISE photometry will be considered in a future work and is essential for determining the performance of this approach when applied to real data at such high SNR values. 

Total exposure times on the order of \qty{5}{\minute} are not unreasonable for a LEO space telescope. 
WISE orbited the Earth at an altitude of $\sim \qty{530}{\kilo \meter}$, completing one orbit every \qty{94}{\minute}\footnote{\url{https://wise2.ipac.caltech.edu/docs/release/prelim/expsup/sec1_1.html}}.

\section{Statistical methods for uncertainty and model preference}\label{sec:apdx:methods}

\subsection{Parameter estimation uncertainty}\label{sec:apdx:methods:uncert}
Sampling distributions for parameter estimation uncertainty results in this work were obtained via Monte Carlo simulations. We simulated a large number (\num{30000}) of experiments consisting of measurement at the central wavelength of each of the WISE IR bandpasses (W1: \qty{3.4}{\micro \meter}, W2: \qty{4.6}{\micro \meter}, W3: \qty{12.0}{\micro \meter}, W4: \qty{22.2}{\micro \meter}) where each of the measurements is sampled from a Gaussian distribution of mean given by the two-component spectral model (eqn. \ref{eqn:2bb}) and a constant relative standard deviation, $\sigma_I/I = \num{1E-3}$ (SNR \num{1000}). That is the experiment is a sample from,
\begin{eqnarray}
  I_i \sim& \mathcal{N}(\mu_i, \sigma_i^2)\\
  \mu_i =& I(u_i,\alpha_0, {f_C}_0, {T_{\text{eff}}}_0, {\Delta T}_0)\\
  \sigma_{i}^2 =&  \mu_i^2 (\sigma_I/I)^2
\end{eqnarray}
Where, $i \in \{ 1,2,3,4\}$ is the observation filter index, $u_i$ is the non-dimensional angular frequency corresponding to the central wavelength of W$i$, and $I$ is given in eqn. \ref{eqn:2bb}. Data from the WISE space telescope and exposure time calculations based on these data has indicated an SNR of at least 1000 is achievable even in the narrower W1 and W2 bands with \qty{5}{\minute} of data collection (\ref{sec:apdx:snr}). While in reality the SNR achieved in each band varies, accurately determining this SNR requires a sophisticated noise model including the effects of temperature-dependent RSO emissions, sky noise, thermal simulation of the space telescope, reflected sunlight subtraction, and detector properties, which was deemed beyond the scope of this conceptual study. 
It is important to highlight that in this work we only consider the spectral irradiance at the central wavelength of a broad-band filter. Therefore, we do not consider the uncertainty in the parameter inference associated to performing the integral over the wavelength of the RSO spectrum convolved with the bandpass throughput for each filter.
This introduces an error which is dependent on the width of the bandpass and the curvature of the spectral irradiance and is an important consideration which will be included in future iterations of the model. 
Quantification of this error is presented in \ref{sec:apdx:broadbands}.

For each of these experiments, we find the maximum likelihood estimate for the parameters of eqn. \ref{eqn:2bb} ($\hat{\alpha}, \hat{f_C}, \hat{T_{\text{eff}}}$, and $\hat{\Delta T}$) and from which $\hat{T_C}$ and $\hat{T_H}$ are also calculated. These maximum likelihood estimates form the sampling distribution from which parameter uncertainties are derived.
As our data are Gaussian this reduces to solving the non-linear least squares problem:
\begin{eqnarray}
  \{ \hat{\alpha}, \hat{f_C}, \hat{T_{\text{eff}}}, \hat{\Delta T} \} = \argmin_{\{ \alpha, f_C, T_{\text{eff}}, \Delta T \} \in D} \sum_i \left(\frac{ I(u_i,\alpha, f_C, T_{\text{eff}}, \Delta T) - I_i}{\sigma_{i}}\right)^2 \label{eqn:fit}
\end{eqnarray}
where $D$ is domain we constrain our estimates to be within and is given by $\hat{\alpha} \in (\num{E-2},\num{E2})$, $\hat{f_C} \in [0,1]$, $\hat{T_{\text{eff}}} \in (\qty{100}{\kelvin}, \qty{700}{\kelvin})$, and $\hat{\Delta T} \in [\qty{0}{\kelvin}, \qty{600}{\kelvin})$.
In the limit that $\Delta T \rightarrow 0$ only one component is needed to describe the data. In this case, the other component does not provide physically meaningful information about the RSO.
To ensure that both components are physically meaningful, we consider only those fits where each model component has a sum of its fluxes across all measurement bands greater than \qty{1}{\percent} of the total flux. That is:
\begin{eqnarray}
    \min \left( \sum_i I_1(u_i, \hat{\alpha} \hat{f_C}, \hat{T_C}), \sum_i I_1(u_i, \hat{\alpha} (1 - \hat{f_C}), \hat{T_H}) \right) > \qty{1}{\percent} \sum_i I_i. \label{eqn:physical}
\end{eqnarray}
This minimization problem (eqn. \ref{eqn:fit}) was solved using the \texttt{curve\_fit} function from \texttt{scipy} to obtain these least squares fit parameters.
As our data are Gaussian, the least squares fit produces the maximum likelihood estimate, however as our data are limited (measurement in only 4 bands) these estimates will, in general, be biased.
Due to this, we quantify the parameter uncertainty by calculating the root mean square error for each parameter. 

\subsection{Model preference}\label{sec:apdx:methods:preference}
This appendix provides the method by which presented model comparison results were obtained.
Given the one-component model is nested within the two-component model, the likelihood ratio test can be used to determine which is preferred. This test, however, returns the preference for given measurement data in each of the sampled wavelengths.
In our analysis, we are not interested in the preference for any specific set of measurements but rather in the proportion of observations for which the two-component model is preferred.
We therefore realize \num{10000} sets of measurements and determine for each which model is preferred. 
We state that the two-component model is preferred over the one-component model when both a) eqn. \ref{eqn:physical} is true for the fit (i.e. the fit is physical) and b) the $|\Delta \chi^2|$ metric suggests the two-component model is preferred at a significance of $\num{0.05}$ (that is $|\Delta \chi^2 | > 5.991$), where
\begin{eqnarray}
  \Delta \chi^2 =& \chi^2_1 - \chi^2_2\\
  \chi^2_1 =& \min_{\{ \alpha, T \} \in D_1} \sum_i \left(\frac{I_1(u_i,\alpha, T) - I_i}{\sigma_{i}}\right)^2\\
  \chi^2_2 =& \min_{\{ \alpha, f_C, T_{\text{eff}}, \Delta T \} \in D} \sum_i \left(\frac{I(u_i,\alpha, f_C, T_{\text{eff}}, \Delta T) - I_i}{\sigma_{i}}\right)^2
\end{eqnarray}
and $D_1$ is the restriction of $D$ to the $\alpha$, $T$ subspace. Again, this minimization problem was solved using the \texttt{curve\_fit} function from \texttt{scipy}.
This preference test is repeated for each ${\Delta T}_0, {f_C}_0$ pairing.

For each realization, measurements of the two-component model in eight filters of Gaussian uncertainty $\sigma_I/I = \num{E-3}$ (SNR 1000) logarithmically spaced in wavelength between the central wavelengths of WISE W1 and W4 were taken. To meaningfully compare the one- and two-component models, a positive number of degrees of freedom in the fit is required. Therefore, we chose to measure in eight filters. Eight filters was also shown above to be a good tradeoff between the number of bands and decrease in measurement uncertainty.

\section{Model uncertainty from central wavelength sampling}\label{sec:apdx:broadbands}

For simplicity, all results presented in this work are based on assuming that the photometric measurement is at a single (central) wavelength, at odds with most imaging systems that would typically use broadbands to collect more photons. Therefore, it is critical to assess whether our results are robust against this assumption. 

An estimate for the introduced error can be made via the Fisher matrix formalism \citep{huterer_course_2023}. The Fisher matrix formalism models the sampling distribution as a multi-variate normal distribution about the true parameter values and is most accurate in the high SNR limit.
An exact error will depend on the specific observation bands, their spectral responses, the spectra of the RSO, and the SNR in each band, however we use the WISE bands and their spectral responses together with a constant SNR, $Q$, as a single point of comparison. 

As our data are Gaussian and independent the Fisher matrix for central wavelength sampling is given by,
\begin{eqnarray}
  \mathcal{F}_{\text{cw}}(\alpha, f_C, T_{\text{eff}}, \Delta T) =& \sum_i \left( \frac{1}{\sigma_{\text{cw}~i}^2} \left. J_I^T J_I\right|_{u_i, \alpha, f_C, T_{\text{eff}}, \Delta T} \right),\\
  J_I =& \begin{bmatrix}
    \diffp{I}{\alpha} & \diffp{I}{f_C} & \diffp{I}{T_{\text{eff}}} & \diffp{I}{\Delta T}.
  \end{bmatrix}
\end{eqnarray}
That is $J_I$ is the Jacobian of $I$ (eqn. \ref{eqn:2bb}) only in the variables corresponding to the parameters of the two-component spectral model ($\alpha, f_C, T_{\text{eff}}$, and $\Delta T$), and $\sigma_{\text{cw}~i} = I(u_i, \alpha, f_C, T_{\text{eff}}, \Delta T)/Q$ is the  uncertainty in the central wavelength sampling of $I$ at $u_i$.

For the WISE bands and spectral responses the measured signal in W$i$ is given by,
\begin{eqnarray}
  S_i = \int_{0}^{\infty} \dif u R_i(u) \frac{I(u)}{u},
\end{eqnarray}
where $R_i$ is the relative spectral response for W$i$ \citep{wright_wide-field_2010}. Likewise, the Fisher matrix for this broadband sampling is given by,
\begin{eqnarray}
  \mathcal{F}_{\text{bb}}(\alpha, f_C, T_{\text{eff}}, \Delta T) =& \sum_i \left( \frac{1}{\sigma_{S_i}^2} \left. J_{S_i}^T J_{S_i}\right|_{\alpha, f_C, T_{\text{eff}}, \Delta T} \right),\\
  J_{S_i} =& \begin{bmatrix}
    \diffp{S_i}{\alpha} & \diffp{S_i}{f_C} & \diffp{S_i}{T_{\text{eff}}} & \diffp{S_i}{\Delta T}
  \end{bmatrix}
\end{eqnarray}
Where $J_{S_i}$ is the Jacobian of $S_i$ and $\sigma_{S_i} = S_i/Q$ is the uncertainty in the measurement of $S_i$. 

Given these Fisher matrices, the Cramer-Rao bound can be applied to give a lower bound on the uncertainty in the sampling distribution of the parameters $\{\hat{\alpha}, \hat{f_C}, \hat{T_{\text{eff}}}, \hat{\Delta T} \}$. That is,
\begin{eqnarray}
  \sigma(\hat{\alpha}) \geq \sqrt{(\mathcal{F}^{-1})_{11}}, \quad
  \sigma(\hat{f_C}) \geq \sqrt{(\mathcal{F}^{-1})_{22}}, \quad
  \sigma(\hat{T_{\text{eff}}}) \geq \sqrt{(\mathcal{F}^{-1})_{33}}, \quad
  \sigma(\hat{\Delta T}) \geq \sqrt{(\mathcal{F}^{-1})_{44}}.
\end{eqnarray}

The ratio of the broadband to central wavelength lower bounds ($\sqrt{(\mathcal{F}_{\text{bb}}^{-1})_{jj}/(\mathcal{F}_{\text{cw}}^{-1})_{jj}}$) gives an indication of the expected relative increase in uncertainty that would be seen with broadband photometry. Evaluating this ratio over the physically relevant parameter space $\alpha_0 \in [\num{E-2},\num{E2}]$, ${f_C}_0 \in [0,1]$, ${T_{\text{eff}}}_0 \in [\qty{200}{\kelvin}, \qty{350}{\kelvin}]$, and ${\Delta T}_0 \in [\qty{10}{\kelvin}, \qty{70}{\kelvin}]$ it is seen that it achieves a maximum of $1.21$ for $\hat{\alpha}$, $1.44$ for $\hat{f_C}$, $1.26$ for $\hat{T_{\text{eff}}}$, and $1.51$ for $\hat{\Delta T}$. Notably for each parameter, this maximum is achieved at the ${T_{\text{eff}}}_0 = \qty{200}{\kelvin}$ boundary of the parameter space with smaller ratios attained throughout the rest of the space. This also indicates that the ratio is likely to increase at cooler ${T_{\text{eff}}}_0$, outside the physically relevant parameter space.
Regardless, a maximum increase in parameter uncertainty by about \qty{50}{\percent} is estimated when using the WISE broadband filters as opposed to central wavelength sampling of these same filters. 

\section{Justification of temperature uncertainty requirements of SDA capabilities}\label{sec:apdx:caps}

To justify and the positions of each of the capabilities illustrated in \cref{fig:fontier}, below we provide a brief argument for a sufficient the temperature uncertainty or resolvable component temperature difference needed to establish each capability.

\textbf{Eclipse:} during eclipse solar irradiance suddenly drops to near-zero and RSOs cool as they continue to radiate the infrared. For planar and low thermal mass components such as satellite solar arrays temperature drops of order \qty{100}{\kelvin} are expected over a \qty{1}{\hour} eclipse period \citep{kim_analytical_2010}. The satellite chassis and other components with greater thermal mass also cool, however at a lesser rate \citep{gilmore_spacecraft_2002}. Therefore temperature measurement to within \qty{10}{\kelvin} is more than sufficient to detect satellites entering eclipse.

\textbf{Seasonal illumination:} due to Earth's axial tilt ($\psi = \qty{23.4}{\degree}$) over the course of a year geostationary RSO will experience a change in beta angle (the angle between the orbital plane and the sun vector) from $-\psi$ to $\psi$. For sun-facing solar arrays with one degree of tracking freedom, this will cause a change in solar irradiance from a maximum of $I_0$ at equinox to $I_0 \cos \psi$ at solstice. Using this range of $\psi$ for evaluating eqn. \ref{eqn:balance} a \qty{7}{\kelvin} difference in the steady state temperatures of the solar arrays are seen due to seasonal illumination.

\textbf{Electric thrusters:} electric thruster consume large amounts of power during operation and obtain temperatures several hundred above ambient spacecraft temperatures. For instance the HERMeS thruster has a nominal operating power of \qty{10}{\kilo\watt} and has surfaces of order \qty{0.1}{\meter} at temperatures ranging \qtyrange{606}{855}{\kelvin} while firing \citep{myers_hall_2016}.

\textbf{Characterization:} as following from our thermal study temperature differences of up \qty{90}{\kelvin} are expected between a box-wing satellite's body and solar arrays when idling. Factors such as power flow from the solar arrays to the satellite body and certain spacecraft geometries and material properties may reduce this temperature difference, and a target of characterizing the presence of a second greybody component with at least a \qty{30}{\kelvin} temperature difference provides a more conservative figure.

\textbf{Power flows:} from our thermal modelling results of a $\approx \qty{1}{\meter}$-sized geostationary box-wing satellite in \cref{fig:steady-state}, a near-constant gradient of $\qty{70}{\kelvin \per \kilo \watt}$ is seen in $\Delta T$ against power flow. At $\Delta T = \qty{30}{\kelvin}$ our results show that $\Delta T$ can be inferred down to $\qty{3.5}{\kelvin}$. Attributing changes in $\Delta T$ to changes in power flow, this uncertainty in $\Delta T$ corresponds to an uncertainty of \qty{50}{\watt} in power flow, or about \qty{5}{\percent} of the maximum expected power flow - an reasonable threshold to claim power flow monitoring capabilities at.

\end{document}